\def\papertype{preprint}

\def\pMS{manuscript}
\def\pPP{preprint}

\def\figfont{\ifx\papertype\pPP \small\sf \fi}

\ifx\papertype\pPP
\makeatletter  
\def\thebibliography{\section*{References\@mkboth
 {REFERENCES}{REFERENCES}}\list
 {[\arabic{enumi}]}{\leftmargin 1em\labelwidth\z@\labelsep\z@\itemindent -1em
 \parsep 0.0ex 
 \usecounter{enumi}}
 \def\newblock{\hskip .11em plus .33em minus -.07em}
 \sloppy\clubpenalty4000\widowpenalty4000
 \sfcode`\.=1000\relax}
\makeatother   
\fi

\ifx\papertype\pMS
\documentstyle[12pt,aasms4,astrobib]{article}
\fi
\ifx\papertype\pPP

\documentstyle[12pt,aaspp4,astrobib]{article}
\fi

\begin{document}

\title{The Hubble Deep Field: Observations, Data Reduction, and Galaxy Photometry}

\author{Robert E. Williams, Brett Blacker, Mark Dickinson,
W. Van Dyke Dixon\altaffilmark{1},
Henry C. Ferguson,
Andrew S. Fruchter,
Mauro Giavalisco\altaffilmark{2},
Ronald L. Gilliland,
Inge Heyer,
Rocio Katsanis,
Zolt Levay,
Ray A. Lucas,
Douglas B. McElroy\altaffilmark{3},
Larry Petro,
Marc Postman
}
\affil{Space Telescope Science Institute, 3700 San Martin Drive\\
Baltimore, MD 21218
}
\author{Hans-Martin Adorf, Richard N. Hook}
\affil{Space Telescope European Coordinating Facility, c/o ESO,
Karl-Schwarzschild-Str. 2, D-85748 Garching, Germany}

\altaffiltext{1}{Current address: Space Sciences Laboratory, University
of California, Berkeley, CA 94720}
\altaffiltext{2}{Hubble Fellow; 
Current address: Carnegie Observatories, 813 Santa Barbara
Street\\ Pasadena, CA 91101}
\altaffiltext{3}{Current address: Jet Propulsion Laboratory, California Institute of Technology,\\
4800 Oak Grove Dr., Pasadena, CA 91109}


\ifx\papertype\pPP \singlespace \fi
\begin{abstract}

        The Hubble Deep Field (HDF) is a Director's Discretionary program on
HST in Cycle 5 to image an undistinguished field at high Galactic latitude in
four passbands as deeply as reasonably possible. These images provide the most
detailed view to date of distant field galaxies and are likely to be important
for a wide range of studies in galaxy evolution and cosmology. In order to
optimize observing in the time available, a field in the northern continuous
viewing zone was selected and images were taken for ten
consecutive days, or approximately 150 orbits.  Shorter 1-2 orbit images were
obtained of the fields immediately adjacent to the primary HDF in order
to facilitate spectroscopic follow-up by ground-based telescopes.  The
observations were made from 18 to 30 December 1995, and both raw and reduced
data have been put in the public domain as a community service. 
        We present a summary of the criteria for selecting the field, the 
rationale behind the filter selection and observing times in each band,
and the strategies for planning the observations to maximize the exposure time
while avoiding earth-scattered light. Data reduction procedures are outlined,
and images of the combined frames in each band are presented. Objects detected
in these images are listed in a catalog with their basic photometric parameters.

\end{abstract}

\ifx\papertype\pMS
\clearpage
\fi

\section{Introduction}\label{secintro}

\ifx\papertype\pPP \singlespace \fi

The HDF program is an outgrowth of previous, highly successful
Hubble Space Telescope imaging projects which have elucidated the 
evolution of galaxies at high redshift.   During Cycles 1 
through 5, a variety of HST General Observer and Guaranteed Time Observer 
programs, as well as the 
Medium Deep Survey (MDS) key project, imaged distant galaxies in 
both cluster and field environments, providing (for the first time) 
kiloparsec--scale morphological data at all redshifts.
The MDS used the
WFPC-1 and WFPC-2 cameras in parallel mode to image random galaxies near the
fields of targeted objects.
Analyzing 144 field galaxies having $I < 22$ from six fields,
\citeN{DWG95} found from visual classification that
early-type spirals, ellipticals, and late-type spirals/irregulars were observed
in roughly equal proportions, with the Sd/Irr's having much higher surface
density than their counterparts at the current epoch.  \citeN{DWOKGR95}
extended this analysis with a similar study of one very deep field 
for which they showed that galaxy counts beyond $I = 22$ continue to
be increasingly dominated by Sd/Irr galaxies.
Combining ground-based redshift information with HST imaging, Lilly and
collaborators obtained B and I images for 32 galaxies from their CFHT survey
($17.5 < I < 22.5$) with known redshifts in the range $0.5 < z < 1.2$ 
\cite{SLCHLT95}.
They found that the observed galaxy morphologies were similar to
those seen locally, but that the B images (rest frame UV) looked far less
regular than observed at longer wavelengths.  In addition, they determined
that the central surface brightnesses of the disks in their sample of late-type
spirals were more than 1.2 magnitude brighter than found locally.  Also, they
found that many of the bluer galaxies were nucleated, and they concluded that
both of these effects must be responsible for much of the observed evolution
of the luminosity function of blue galaxies.

   Other HST programs targeted galaxies with known redshifts based upon
their membership in clusters that had been studied from the ground, e.g.,
0939+4713 (Dressler et al. 1994a,b) \nocite{DOBG94,DOSL94}
and the cluster(s) associated with the radio
galaxy 3C 324 at z = 1.21 \cite{Dickinson95Gal}. 
Both of these programs demonstrated the ability of the refurbished HST
to resolve galaxy structure at moderate to high redshift in a way that
made morphological classification and a quantitative study of various
parameters possible.  Cluster 0939+4713 does not look entirely unlike
nearby clusters insofar as it is populated largely by 
spiral and elliptical galaxies.
However the disk systems are bluer and more numerous than spiral
galaxies in the cores of clusters today, and often show signs
of disturbance and tidal interactions.  Evidently, these spirals
are responsible for the rapidly evolving blue galaxy population first
noted in distant clusters by Butcher \& Oemler (1978; 1984).
\nocite{BO78,BO84}
Looking back to $z = 1.21$, the cluster associated with 3C~324
includes apparently normal, mature E/S0s, but readily recognizable
spiral galaxies appear to be rare, and a large number of irregular,
amorphous objects are present (Dickinson 1995a,b).
\nocite{Dickinson95Gal,Dickinson95Fresh}
Since the first servicing mission, HST has imaged a number of other distant galaxies
at still higher redshifts 
of $1 < z < 3.5$ (c.f. \citeNP{CHS95}; \citeNP{GSM96})

While much of the information available in these images remains to be interpreted,
two things have become clear.  First, HST can indeed resolve galaxy-sized
systems out to high redshift. Second, the Universe at high redshift looks 
rather different than it does at the current epoch.  The fact that
HST can image galaxies back at epochs when they were apparently
forming and evolving rapidly is of fundamental importance to our
understanding of galaxy evolution, and it is imperative that this
capability be fully exploited. 
Based on the current excellent
performance of the telescope, a decision was made to devote a
substantial fraction of the Director's Discretionary time in Cycle 5
to the study of distant galaxies.  A special Institute Advisory
Committee was convened which recommended to the Director that deep 
imaging of one `typical' field at high galactic latitude be done with
the Wide-Field Planetary Camera 2 (WFPC-2)
in several filters, and that the data be made available
immediately to the astronomical community for study.  Following this recommendation
a working group was formed to develop and carry out the project.

It is not our purpose here to interpret the data, but rather to present
the images and source catalogs, along with the necessary background to
facilitate the use of the HDF in studies of galaxy evolution.  As part
of the background, we describe the criteria for selecting the field,
the scientific rationale for the selection of filters, technical
aspects of planning the observations, and details of data reduction and
calibration. The images are presented and discussed in
\S\ref{secimages}, and the source catalogs in \S\ref{seccatalogs}.  In
view of the wide interest in these observations, we have
tried to provide a useful reference work, being reasonably
comprehensive in describing the data reduction, and including in
printed form the most important photometric parameters for the detected
sources.  Nevertheless many of the parameters for the individual
galaxies (e.g. S/N in the different bands, higher-order image moments,
star/galaxy classifications) have been left out of the printed catalog.
We have opted to emphasize black \& white over color images to provide
as much detail as possible in the limited dynamic range of the printed
page. The full catalog and color images are maintained on the STScI
world-wide web site at http://www.stsci.edu.

\section{Field Selection}\label{secfield}
\subsection{Primary Field }\label{sechdffield}

One of the suggestions to the Advisory Committee by the Institute 
was that a field 
in one of the continuous viewing zones (CVZ) be considered, because
the observing efficiency there could be up to a factor of two higher
than other locations on the sky.
The working group focused its attention on the 
northern CVZ, thereby constraining the HDF location to a narrow 
declination range centered around $+62^{\circ}$. 
In addition to being in the CVZ, the candidate field would have to have
low optical extinction ($E(B-V) < 0.01$ mag), low HI column density
($< 2.5 \times 10^{20}$ cm$^{-2}$), and
low FIR ($\lambda = 100\mu \rm m$) cirrus emission. Furthermore,
to facilitate faint object studies at many wavelengths, 
the HDF field would need to avoid known bright sources
in the x-ray, UV, optical, IR, and radio passbands. 
This latter criterion, therefore, excluded areas with known nearby ($z
< 0.3$) galaxy clusters. Visual inspection of IPAC's co-added 100$\mu \rm m$
IRAS maps was used to select regions with no significant Galactic
cirrus features.  An initial sample of about 20 possible HDF regions
was narrowed to 3 optimal candidates, all within the Ursa Major
region.  Quick VLA snapshots at 3.6 and 21 cm by \citeN{Kellermann95} reduced
the choice to two fields due to the presence of a 68 mJy source in the
center of one of the fields. Initial optical selection was based on the
digitized Palomar sky survey.  Eisenhardt \citeyear{Eisenhardt95} kindly provided KPNO
4m R-band CCD images ($2 \times 300$ second exposures) as further verification
that the fields were typical in terms of source counts, and were not
affected by scattered light or diffraction spikes from bright stars
outside the field

A search of the ROSAT data archive \cite{Petre95} in the vicinity
of the two remaining field candidates 
placed a conservative upper limit on the flux from any source
of $6 \times 10^{-14}$ erg cm$^{-2}$ sec$^{-1}$ in the 0.1-2.4 keV band.

In June 1995, HST acquired a single orbit F606W 
observation of each of the two fields to verify guide star acquisition.
In order to be conservative in safeguarding the entire
sequence of HDF observations, we required an independent pair of back-up guide
stars, which are scarce at this high Galactic latitude.  The decision
between the two remaining candidate fields was thus based on guide star
availability. The location and characteristics of the resulting HDF 
field are given in Table \ref{tabfield}, and the KPNO R-band image of the field
telescope is shown in Fig. \ref{figfield}
with the WFPC-2 field of view superposed. 

\subsection{Flanking Fields}\label{secflanking}

During the HDF observations, 10 orbits were devoted to short
exposures of eight ``flanking'' fields adjacent to the main survey region. 
These fields were arranged in a roughly square pattern surrounding
the central HDF, as shown in Fig. \ref{figfield}.
All exposures in the flanking fields were taken at the same orientation as the
central field. 
The coordinates for these fields are given in Table \ref{tabflanking}. 

\section{Filter Selection}\label{secfilters}

The selection of filters for the HDF observations
represents a balance between the desire for depth and the desire for
color information and practical considerations involving scattered
earth light. 

The HDF observations were not aimed at answering one specific question,
but are rather intended to be of general use for constraining models of
galaxy evolution and cosmology. There was thus no single, well-defined
criterion that could be used to develop the optimum observing strategy.
For galaxies that are well resolved, color gradients and the dependence
of morphology on wavelength are of interest. If such studies
were the sole aim of the observations, the best strategy might be
to opt for the highest possible S/N in two widely separated
bandpasses.  On the other hand, for the more numerous faint and barely
resolved galaxies, it is the statistical distributions of color vs.
magnitude, color vs. angular size, etc. that are of interest. These
provide information on both the redshift distribution and the stellar
populations of the galaxies. For galaxies fainter than spectroscopic
limits, it is important to have at least two colors. A single color (or
``spectral index'') is less useful because there is no way to separate
effects of the intrinsic rest-frame spectral properties from the effect
of the $k$-correction. This argues for observations through at least
three filters, preferably with the highest possible efficiency and with
minimal overlap in bandpasses. While other options offer slightly
better photometric accuracy or slightly less bandpass overlap, the
combination of F450W, F606W, and F814W provides a very efficient way to
cover the optical portion of the WFPC-2 bandpass (Fig. \ref{figfilters}).
The addition of observations through the F300W filter greatly improves the leverage for
statistical redshifts. These observations cannot reach the limiting depth of the other
three filters because of the low detector quantum efficiency (QE) at
3000 {\AA}. However, because of the low QE, the F300W data are largely
read-noise limited, whereas the other bands are more nearly sky-noise
limited. Thus the the F300W observations can make use of the orbital
``bright time'' (see below) that is not particularly useful for the
other filters.

With this four-filter strategy the HDF reaches depths roughly
three magnitudes fainter than the deepest ground based images in the
red bands, two magnitudes deeper in the B band, and one magnitude
deeper in the U band. The 80\% completeness limit of current deep
spectroscopic surveys is $B_{AB} \sim 24$
\footnote{Most magnitudes in this paper are
expressed in the $AB$ system (Oke 1974), where
$m = -2.5 \log f_\nu -48.60$.} 
\nocite{Oke74} 
and $I_{AB} \sim 22.5$
\cite{GECBAT95,CFRS1}. With Keck and other large
aperture telescopes, these limits may be pushed one or even two magnitudes
fainter.  Even then, the faintest two or three magnitudes of the HDF survey
are beyond current spectroscopic limits.  Hence, for at least the next few
years, the primary information on the redshift distribution of these
very faint galaxies will come from statistical analysis of the color
distribution. In the individual F450W, F606W, and F814W bands, the HDF
observing strategy provides only a modest improvement over the deepest
existing HST images. However, having four passbands provides essential
information not available from previous surveys.

Part of the scientific interest in F300W stems from its utility in
searches for very high-redshift ($z>3$) galaxies (\citeNP{GTM90,SH92},\citeyearNP{SH93}) The
intrinsic 912{\AA} Lyman break in galaxies, combined with the
increasing opacity of the intergalactic medium at high redshifts
produces a distinctive feature in the spectra of high-redshift galaxies
\cite{YP94,Madau95}. Specific color selection
criteria for the HDF bandpasses are described by \citeN{MFDGSF96}.
The F300W filter has a small but significant ``redleak'' at 8000{\AA}.
The transmission curve of this redleak is similar to that of the
F814W filter, but the throughput is three orders of magnitude smaller.
Thus, a 20th magnitude galaxy with no intrinsic flux in the F300W bandpass
will appear as an F300W source of AB magnitude $\sim 27.7$. Most of
the galaxies detected in the HDF have magnitudes fainter than $I_{814} = 23$,
and would thus be below the detection limits of the F300W images unless 
they have intrinsic emission near 3000{\AA}.

\subsection{Scattered Light and Observation Scheduling}

The choice of the F300W passband was motivated partly 
by the desire to provide a measure
of the surface density of galaxies above redshift $z=3$
and partly by the desire to
find an observing strategy that avoids contamination from stray earth
light.  While the zodiacal-light background in the HDF field is
extremely low in December, scattered earth light during portions of the
CVZ was expected to dominate the background in F450W, F606W, and F814W.
The scattered earthshine in the F300W filter is more than 15 times fainter
than in the other three filters due to absorption of the solar flux by
ozone in that band.
HST can in principle observe to within 15.5 degrees of the bright earth
limb and 7.6 degrees of the dark limb (these limits are imposed by the
Fine Guidance Sensors).  However, the effects of scattered light can be
detected in archival WFPC-2 images at angles to the bright limb as great as
40 degrees.  The amount of scattered light in the HST focal plane varies both with limb angle
and the total brightness of the earth. The observations were carefully scheduled to make use of the
dark portion of the orbit for observations through the 
F450W, F606W, and F814W filters, and the bright portion for F300W
observations. 

Efforts over the last two years to understand
the sources of background have resulted in the development of a program
(SEAM; \citeNP{BPEK96}) that successfully models the WFPC-2 background as a function of
wavelength and viewing direction. SEAM has been used to predict the
background in various filters considered for the HDF.
Figure \ref{figbackground}
shows examples of the predicted background
as a function of time in several orbits spaced throughout the
program. Curves such as these were used to plan the start and stop
times of each observation to maximize observing time, while minimizing
the total background. To ensure that the timing was correct, the
start/stop times were adjusted using the latest available orbit
predictions as a last step before the final schedule was frozen
several weeks prior to the observations. 

It should be emphasized that without this careful scheduling
of the observations, the extra observing time gained by selecting
a field in the CVZ would have been largely wasted. While such
micro scheduling is not something that is routinely done for HST
observations, it is clearly something that is desirable to implement
in a more automated manner for observations in which sky background
is the limiting factor in obtaining the best signal-to-noise ratio.

Another decision made in the planning stage was to attempt to obtain
observations at nine separate pointing positions through each filter.
The 0.1$^{\prime\prime}$ pixels of the three wide-field (WF) cameras 
in the WFPC-2 instrument undersample the point spread function of the
telescope, while the planetary camera (PC) provides more nearly optimal
sampling. The motions of the telescope were 
laid out as non-integer multiples of a WF pixel.
This ``dithering'' reduces the photometric errors due to flatfielding
uncertainties and also allows reconstruction of a higher-resolution
image because sources are sampled in different portions of a pixel at
each dither position. The price for these improvements is increased
complexity in the data reduction phase. The images must be aligned and
resampled to the same pixel scale, correcting for the geometric
distortions introduced by the camera optics.  We chose to observe at
nine dither positions spanning a range of 2.6 arcseconds that would in
addition map to nine independent positions within the (WF) pixel scale
of 0.1 arcseconds (Fig. \ref{figdither}). Two unplanned dither
positions with offsets of $\sim 1^{\prime\prime}$ and a rotation of 4.3
arcminutes resulted from an 11 hour period in which one of the Fine
Guidance Sensors locked up on a secondary extremum of the ``S-curve.''
We tried to obtain at least five exposures per filter per dither position that
were near enough in time that slow drifts in the telescope pointing
(which had been seen in previous CVZ observations) would not complicate
the cosmic-ray rejection process.

\section{Data Reduction}
\subsection{Pipeline}

The HDF data were reduced three times. Version 1 reflects
processing up to January 15, 1996. Version 2 was released February
29, 1996 and is used for this paper. At the time of writing,
version 3 was still a month or two from completion. 

The HDF data were reduced using the same software as the standard HST
pipeline calibration, the STSDAS task calwp2, but with different
calibration files and a slightly modified treatment of the darktime.
The calibration procedure is fully documented in the HST Data Handbook
\cite{data_handbook95}.
For the HDF, several of the calibration files were
improved by combining a number of individual calibration frames to
produce ``super'' calibration frames. We discuss these improved
calibration files below. Versions 1 and 2 used the same calibration
files. These are being modified for version 3, for reasons that
will be outlined below.

{\it The superbias frame.} To the charge accumulated in each pixel of
the WFPC-2 CCDs is added a bias value, designed to keep the output of
the analog-to-digital (A/D) conversion consistently above zero. The
value of this bias can vary slightly with position across the chip. A bias
reference file is therefore subtracted from the data to remove any
position-dependent bias pattern. The HDF used a superbias frame
constructed from 160 individual bias frames. These frames, in
uncalibrated format, were retrieved from the archive in sets of 40 and
calibrated (A/D conversion and overscan subtraction) in the standard
way. Each set of frames was then combined into a single image with cosmic
rays removed. The resulting 40-frame bias images were used to create
superdark images (see below) for each epoch. 
Because of the timing in the WFPC-2 electronics, bias frames are not
really zero-length exposures, but have 43.6 seconds (plus the 
readout time for each chip) of exposure time. To remove this, 
the superdarks were 
normalized to exposure times of 43.6 s and subtracted from the
superbiases. The dark-subtracted bias frames were then averaged to
produce the final superbias frame.
To use the
file correctly, the DARKTIME header field in the frame to be calibrated is
updated to the value
\begin{equation}
DARKTIME = 60 \times int((t+16.4)/60.) + 43.6,
\end{equation}
where $t$ is the integration time of the image in seconds.
The standard calwp2 data reduction software
will then scale the superdark appropriately, and the dark
current remaining in the bias frame will be removed in the superbias
subtraction. The version 3 data reduction will use bias frames taken
both before and after the observations so any systematic drifts with
time will be better averaged out. 

{\it The superdark and delta-superdark frames.} The dark subtraction is
intended to remove spatial structure present in the thermally-induced
dark current. In practice, most of this structure is in the form of
individual ``warm'' pixels that are between $2 \sigma$ and
$5 \sigma$ deviant from the mean, but are relatively stable over long
periods of time.  The superdark frame used in the HDF is derived from
eight sets of 30 individual calibrated dark frames (1800 second exposures
taken with the shutter closed). The frames in each
set were combined and cosmic rays removed, then the eight sets were
combined into a single file. The resulting image was normalized to a
darktime of 1.0 second.

A significant fraction of the counts seen in WFPC-2 ``dark'' frames
arise from cosmic-ray glow in the MgF$_2$ faceplates in front of the
CCD's. This glow varies with the cosmic ray rate and for version 3 will
be subtracted independently of the thermal dark current in the CCD
detector. For versions 1 and 2, variations in the amplitude of the
MgF$_2$ glow were ignored (i.e.  the superdark was scaled by exposure
time and no correction was made for the cosmic ray rate).  The result
is that individual images, after flatfielding, typically show curvature
in the background of a few percent within 200 pixels of the edge of the
chip. These variations in background average out to less than 1\% when
multiple frames are stacked together.

To flag the hot pixels present at the start of the HDF observations,
a delta-superdark frame was created. To do this, 11 dark frames 
(total exposure time 19000 seconds) taken near the start
of the HDF observations were averaged together. The superdark
frame was subtracted and pixels more than $5\sigma$ deviant from the
mean were identified.
(Note that ``cold'' pixels --- i.e. those less than $-5\sigma$ deviant --- were
also identified.)
These pixels were added to the superdark frame
(so that they would subtract, to first order, from the individual observations),
and were also flagged in the data quality file, ensuring that they
would be ultimately rejected in the final combinations of the images
at the different dither positions. 
The delta-superdark was the same
for versions 1 and 2, but is being remade for version 3.

{\it The flat-field frames.} The flat-field frames developed for
the HDF have since become the default flats for the HST pipeline
calibration. These files are versions of the original Investigation
Definition Team (IDT) flats with an improved on-orbit illumination
correction applied to scales greater than seven pixels. On the very
largest scales, the chip-to-chip normalization of the IDT flats have
been preserved. Tests of independent sets of flat-field images suggest
that the uncertainties are less 2 \%. More information on these 
files can be found in the WFPC-2 Cycle 4 Calibration Summary \cite{BCB95}.
For version 3 of the HDF data reduction, more recent versions of the
flat-field calibration files are being used.

\subsection {Cosmic Ray Rejection and Initial Image Stacking} \label{secstacking}

The preceding steps in the data reduction were identical for versions 1
and 2 of the HDF data. The versions diverge at the image combination
step in two ways: (1) the individual images were weighted differently
in the two versions and (2) care was taken in version 2 to remove
satellite and space-debris trails from the images. 

Cosmic ray (CR) identification was carried out using the stsdas task
{\it xcrrej} (a modified version of the standard STSDAS {\it crrej} task).
This  program  is  a more  sophisticated  implementation of the
simple idea of averaging several images of a target while removing
cosmic  ray  events. The process begins by computing sky levels 
from the histogram of pixel intensities, estimating the best value by
parabolic interpolation of the peak and the two adjacent points in the
histogram.  These sky levels are subtracted and the images are
renormalized to the average exposure time to allow CR rejection on
frames with differing backgrounds and exposure times. The program then
computes the expected RMS deviation in each pixel based on the minimum
observed value in that pixel and a model for the noise that includes
Poisson noise, read noise, and a noise proportional to the counts (to
account for variations in the peak brightness of bright sources, e.g.
due to pointing jitter).  In the first pass values more than $6\sigma$
above minimum are rejected. In the second pass, the average in each
pixel is used as the expectation value, and values more than $5\sigma$
above and below the average are rejected.  In the third pass the
average is again used and values more than $4\sigma$ above and below are
rejected.

Typical cosmic rays seen in WFPC-2 are not single-pixel events.  To
remove the wings of cosmic rays, pixels adjacent to rejected pixels are
rejected if they are deviant by one-half the sigma-threshold applied to
the initial rejected pixel.

In version 1 of the HDF processed data, the images were combined with
weights proportional to exposure time. This is nearly optimal for
F606W, and F814W, but not for the F450W and F300W images, which
have a significant read-noise contribution.
For version 2, the images were combined with weights proportional to 
the inverse-variance at the mean background level. The variance, in
electrons, at the sky level is computed from the noise model:
\begin{equation}
\sigma^2 = b t + d t + r^2
\end{equation}
where $t$ is the exposure time, $b$ is the sky background rate, 
$d$ is the dark current, and $r$ is the read noise. 
Pixels affected by cosmic rays were not used in the combination
and were set to zero in the output inverse-variance image.
These output inverse-variance images were used to compute the
weights for the final combined images as described in \S\ref{secdrizzle} below.

\subsection {Hot Pixel Removal}

The number of hot pixels is minimized by using the delta-superdark frame
in the HDF pipeline; however, several thousand remain in each frame
after pipeline processing.

``Growing pixels'' are hot pixels that have appeared or disappeared
since the delta-superdark. They were identified by combining sets of
images taken at different dither positions, subtracting a $5 \times 5$ pixel
median filtered image to remove residual objects, and identifying
pixels that deviate from 0 by more than than 5 $\sigma$.  A
growing-pixel mask was created for each day using about 48 hours of
data roughly centered at noon UT. The masks for each day were logically
OR'ed together and OR'ed with the static mask (the list of constant
WFPC-2 bad pixels) and the delta-superdark data quality file (the list
of pixels that became hot between the superdark and the
delta-superdark). In other words, a pixel flagged as suspect in any one 
of these files was marked in the final data quality file, and ultimately
rejected in the final image combination.

This procedure reduces the number of hot pixels to about a dozen per
chip. These are identified with a program that flags individual pixels
more than 6 $\sigma$ deviant from their neighbors in a $3 \times 3$
box. These pixels are OR'ed into the mask to produce one final mask
that flags all pixels that were suspect at any time during the HDF.
Altogether roughly 12000 pixels per chip were masked. Because there are
eight or nine dither positions per filter, this conservative approach
to masking hot pixels has only a modest effect on the final
signal-to-noise ratio, reducing it by about 10\% in the pixels that were
flagged, and this for only about 2\% of the total imaging pixels.

\subsection {Scattered Light Removal}
The HDF observations were carefully scheduled to minimize the effect
of scattered light. Nevertheless, scattered light from the bright earth 
produces a visible X pattern in about 25\% of the HDF data frames. 
The X pattern is caused by shadowing of the scattered light by the
mirror supports in the WFPC-2 relay optics \cite{BRR95}.
Images with a mean background more than a few times higher than the mean
for a given bandpass were excluded from the processing. 
However, most of the affected frames had only a small amount 
of scattered light, and the S/N was not significantly degraded. 
These images were used in version 2, and steps were taken to remove the
scattered light before combining them with the unaffected frames. 
Briefly, the scattered light subtraction was done by registering the bright
and dark images, subtracting the dark from the bright to remove the sources,
smoothing to remove any residual sources left over after subtraction, and
subtracting this smoothed ``sky'' image from the bright frame. This 
procedure removed the X pattern to within a few percent. 

\subsection {Image Registration and Geometric Distortion Correction}

The HDF observations were carried out with nine to eleven different
pointing positions per filter, spanning a range of roughly 2.6 arcseconds
(Fig \ref{figdither}).
After having been fed through the pipeline, the registration of the
individual images was checked by comparing the positions of several
bright sources. This revealed two unexpected large (0.98 arcsec)
shifts, but otherwise suggested that the images at each dither
position were registered to within the errors of such a comparison.
The images with the large shifts have been treated as separate dither positions.

The roughly five images per dither position were cosmic-ray rejected and
stacked into a single image.
In order to measure the shifts and rotations between these stacks,
these images were first geometrically undistorted and resampled using
the ``drizzle" task which is described below.  The drizzle program, in
turn, removes geometric distortion using the polynomial solution
described by \citeN{TVEM95p379}.
The shifts and rotations between the dither
stacks were then measured
using a cross-correlation-based technique on each of the three
wide-field chips and averaging the results, again using the relative
orientations and positions of the chips measured by \citeN{TVEM95p379}.

Except for the two unexpected
dither settings with a rotation of 4.3 arcmin, all other exposures
align perfectly to an accuracy of less than 0.1 arcmin rotation, the
estimated measurement error.  Estimated errors in the shifts are a few
hundredths of a pixel for F450W, F606W, and F814W.
At the time of this writing, the shifts and rotations have not been checked
to the ultimate possible precision in each of the input frames. There is
thus some possibility of improving the resolution and shape of the final
PSF with a more careful combination of the images. However, we anticipate
such processing will produce only a few percent gain in S/N at the faintest
levels, and a modest additional gain in the ability to distinguish stars
from galaxies.
 
Once the shifts and rotations were measured, the images were aligned
and combined using the ``drizzle'' algorithm,
which corrects for geometric distortion and produces an output image
that is sampled on a smaller pixel scale than the input images.
The final pixel scale chosen for both the PC and WF chips
was 0.4 of the original WF pixel or, using the astrometric
solution of \citeN{TVEM95p379} for WF2, 0.03985 arcsec.
 
\subsection {Image Combination}\label{secdrizzle}

The ``drizzle'' algorithm 
was developed for the combination of multiple stacks of dithered, geometrically
distorted, undersampled image frames. It was used to produce the
combined output images of the HDF project from the flat-fielded, cosmic
ray and hot pixel cleaned stacked frames corresponding to the different
dither positions.  The algorithm, which is more formally known as
variable-pixel linear reconstruction, will be described in 
detail in another article \cite{FH96}.  Here 
an overview is provided for those interested in understanding the specific processing
of the HDF.

The algorithm is conceptually simple. Pixels in the original ``input'' images 
are mapped into pixels in the subsampled ``output'' images, taking into account
shifts and rotations between images and the optical distortion of the camera.
The final image is built up by averaging together enough
images taken at different positions that non-uniformities in exposure time
from pixel to pixel in the output image become inconsequential (and are in any
case recorded in the output variance map). 

If there are enough input images, a simple way to improve the resolution of the
output image is to make the area or ``footprint'' of each pixel
in the input frame smaller than the physical pixel size, before mapping it
onto the output image.  To carry forward the analogy, these 
shrunken pixels, or ``drops'', rain down upon the subsampled output
image.  In the case of the HDF, the drops had
linear dimensions one-half that of the input pixel --- slightly larger than
the dimensions of the output subsampled pixels. The flux in each drop is
divided up among the overlapping output pixels in proportion to their
areas of overlap. In the code this is done by breaking the drops into 
$N \times N$ ``droplets" and distributing their flux among the output
pixels. The version 1 and version 2 images were made with $N = 12$. For version
3, the code has been revised to calculate the overlap exactly.

The position of each droplet in the undistorted, subsampled output image
is computed and its value is averaged with the
previous estimate of that pixel value.  This is a
weighted average which uses
the assigned weight for the droplet 
and a weight for the previous
estimate which is the sum of the weights of all droplets previously
averaged into the output pixel.
The image value and weight of the output pixel are updated before
considering the next droplet.   Thus, if a particular drop with value
$i_{xy}$ and weight $w_{xy}$ is to be added to an image with value
$I_{xy}$ and weight $W_{xy}$, the resulting value of the image $I'_{xy}$
and weight $W'_{xy}$ is

\begin{eqnarray}
I'_{xy} &=& \frac{i_{xy}w_{xy} + I_{xy}W_{xy}}{w_{xy} + W_{xy}} \\
W'_{xy}  &=& w_{xy} + W_{xy}
\end{eqnarray}

This procedure is performed for each of the $N^2$ droplets in each of the
pixels of the input image.
In order to preserve resolution, a drop size is used which is smaller
than the input pixel size. 
A particular output pixel may receive no droplets
when drizzling an individual input frame. In Figure \ref{figdrizzle} the top left
output pixel represents such a situation. These ``zero valleys" are not a
concern as long as there are enough input frames with different sub-pixel
dither positions to fill in the image. It is, in the end, the 
placement of the
dither positions which determines how small the drop size can be. 

This scheme was developed in part as a quick means of implementing an 
area-weighted interpolation for distorted pixels, similar
to the interpolation scheme proposed by \citeN{TVEM95p379}.  
The weight of a droplet from a particlar pixel is just $1/N^2$ times
the weight of the pixel, which was computed as described
in \S\ref{secstacking}. Therefore, to the
extent that $N$ is large, the weight of a particular input pixel in
a final output pixel is the fraction of the input pixel overlapping
with the output pixel times the input pixel weight.

This algorithm has the following characteristics. 
(1) It preserves both surface and absolute photometry;  
     flux density can be measured using an aperture whose size is independent of 
     position on the chip. 
(2) It handles missing data due to cosmic ray hits and hot
pixels. 
(3) It uses a linear weighting scheme which is statistically optimum
   when inverse variance maps
   are used as weights. These weights may vary spatially
   to accommodate changing signal-to-noise ratios across
   input frames (e.g. due to variable scattered light). This spatial variation
   was not included for versions 1 and 2 of the data reduction.
(4) It produces an inverse variance map (the weight
   map) along with the combined output frame. 
(5) It preserves resolution. 
(6) It largely eliminates the distortion of absolute photometry produced
   by the flat-fielding of the geometrically distorted images. In an
   uncorrected image the total photometry of sources near the corners of
   the chip is about
   3.5\% brighter than an equivalent source at the center of the chip
   \cite{Biretta95p257}.

Drizzling does however produce small artifacts in the final
image.  The interpolation scheme used 
produces an output image that is optimal
 for aperture photometry when the output image is
weighted by the inverse variance. However,   
it is not
optimal if one is interested primarily in producing a smooth
point spread function (PSF). Simulations of the dithering pattern and
drizzle parameters used in the HDF show that even in the absence of
Poisson errors, drizzling produces a PSF whose FWHM varies by $\pm 5$\%
and whose shape can show noticeable high-frequency ``noise".  Both of
these effects depend upon the pixel phase of the star relative to the
dither positions, and thus are hard to predict for any individual
object. (This directly limits the possibility of applying
deconvolution, a process that amplifies high frequencies, to the
drizzled data.) In Fig. \ref{figpsf} we show an example of a PSF from
a star in the HDF. Variation about a Gaussian fit is a natural feature
of an image convolved with large square pixels. However, because the
shifts between images in the HDF do not uniformly
sample the plane, the interpolation scheme used in drizzling makes this
high-frequency scatter even more apparent.

The observed variations could have been reduced by using a larger final
pixel size. However, doing so would have meant either using a different pixel
scale for the PC than for the WF or suffering even further degradation of the
PC resolution. While the high frequency scatter will affect attempts to fit
the PSF, it does not significantly affect aperture photometry. 
By a radius of two WF pixels (five pixels in the drizzled HDF images), 
which is an aperture frequently favored
by those doing HST photometry in crowded regions, the scatter is essentially
gone. Furthermore, the scatter is seen only where the variations in surface
brightness of an object are so rapid that they are undersampled in the
original image (point sources, for example). 
Those for whom the scatter is a problem may wish to
convolve the image with a narrow Gaussian, and thus trade a little 
resolution for a smoother PSF. 

Drizzling also
causes the noise in one pixel to be correlated with the noise in an adjacent
one, because a single pixel from an input image typically affects the
value in several output pixels, even though most of the power often goes into
a single output pixel.   The amplitude of this effect is determined by the
footprint, or drop size, of the input pixels and by the positions of the
sub-pixel dither points.  If the data had been combined using the
shift-and-add technique, which has a pixel footprint four times larger
than that used in the HDF, this effect would have been substantially greater.

The drizzling parameters used in the HDF produce negligible
correlations between pixels which are not adjacent.  The measured
$3\times 3$ correlation matrix of adjacent pixels in the F606W frames
is shown in Table \ref{tabmatrix}.  It includes any correlation in the
sky due to inexact flat fielding or true variations in the sky.
In the end, this provides a better idea of how sky noise will be
reduced as the size of an aperture is increased.  The correlation matrix
varies both locally and globally due to the changes in sub-pixel
placement caused by geometric distortion.  The above matrix appears to
provide a good indication of the overall noise statistics of the
image.  Simulated images with the same small-scale noise correlations
as the HDF drizzled images can be created by convolving an image which
has independent noise in each pixel with the matrix shown in Table
\ref{tabnoisekernel}.
The resulting image will have an RMS pixel noise identical to the
input image, but because of the correlation introduced between
neighboring pixels, the expected standard deviation of an $N \times N$ box of
pixels (where N is much larger than 1) is 1/N times the sum of the
above matrix elements, or about 1.9/N.  This procedure only simulates
the correlation of pixels with equal intrinsic noise; a more complete
simulation would need also to allow for the small pixel-to-pixel
variations in the standard deviation of the noise.

\section{The Images}\label{secimages}

Figure \ref{figcolor}
shows a color composite of the HDF full field from the
F450W, F606W, and F814W images. This image was produced from the initial
version 1 data reduction, and has slightly lower S/N than the version
2 images. Nevertheless it reveals the striking variety of colors and 
morphologies of the distant galaxies visible in the field.

   The final, combined images of the HDF which result from the drizzling
process in each of the four filters are presented in Figs. 
\ref{figF300_1} -- \ref{figF814_4}.  In
each figure we display the co-added images from each of the four individual
WFPC-2 CCD's.  The total number of exposures and combined exposure time for
each of the images is given in Table \ref{tabexptimes}.  Also shown are
the sky levels, in units of data numbers (DN) per pixel (one DN corresponds
to approximately 7 detected electrons).  In the figures, the axes show pixel positions
in the same units as the catalog listings. The scale at the bottom shows count levels
in DN/1000 sec corresponding to different intensities on the final print.
The images for the different bands have been scaled so that the maximum gray level
corresponds to a constant AB magnitude of 20.35 magnitudes per square arcsec.
The minimum gray level is set at $-3$ times the RMS of the sky level. For the dark
gray levels in the images, therefore, a galaxy with a flat spectrum in $f_\nu$ will
have a constant brightness in the different bands. However for fainter galaxies, the 
transfer functions diverge for the different bandpasses.

It is clear from visual inspection of
the images that detection differ in each filter due the
variations in their background noise levels.  The higher
background noise of the F300W images, for which the individual exposures
were all read-noise limited, is clearly evident.  Objects that are
present above the background noise display a wide variety of morphologies.
At brightness levels well above the detection limit there are relatively few
stars compared to the number of galaxies.  The galaxies have a wide
distribution of brightnesses, sizes, and shapes, and the brightest galaxies
appear to correspond to the normal Hubble types.  Most of the fainter sources
also appear to be galaxies, although this must be established carefully
because many of them are only marginally spatially resolved, but their
morphologies are frequently chaotic and asymmetric.  Not surprisingly, the fainter
sources have a more compact appearance.  To what extent this is an artifact of the
$(1 + z)^4$ diminution in surface brightness produced by increasing distance
and the changing metric needs to be ascertained.  Some of the fainter objects
are undoubtedly bright H II regions and massive star complexes in galaxies,
which appear above the background threshold while the remaining lower
surface brightness regions of the galaxy do not. 
   A cursory study of the images does not reveal any obvious heretofore
unobserved class of objects compared to earlier HST images of moderately
distant clusters such as 0939+4713 \cite{DOSL94} and those associated
with 3C 324 \cite{Dickinson3C324}.  There do appear to be a number of the
linear structures having the sizes and luminosities of galaxies that have been
noted by others (e.g. \citeNP{CHS95}).  One of the most distinctive galaxies present
in the field is a relatively large, extended, low-surface-brightness galaxy near
the NW edge of WF3. This galaxy appears to be relatively nearby, and it
might be an outlying galaxy member of one of the Ursa Major clusters.
Color differences among the various galaxies are notable in that
some of them appear much more prominent relative to neighboring galaxies in
one of the filters than in others.  Also, different colors in distinct regions 
of the same galaxy, such as the H II regions, are evident in many of the objects.

\section{Source Detection and Photometry}\label{seccatalogs}

\subsection{Overview}

The photometric parameters of very faint galaxies are difficult to
measure and difficult to interpret in a straightforward way. Systematic
errors can arise at every phase of the analysis. During the detection
phase, spurious ``galaxies'' can be identified from noise peaks if the
noise properties of the data are not well quantified; galaxies with
unusual surface brightness profiles can be missed if their sizes are
not well matched by the convolution kernel used to smooth the data.  In
the photometry phase, systematic errors can arise in converting from
isophotal or aperture magnitudes to total magnitudes, in measuring
colors based on isophotes defined in only one band, and in measuring
moments or radii of galaxies using only light from within a certain
radius.  In the source counting phase, galaxies may be overcounted if
substructures are included as separate objects, or undercounted if
overlapping distinct objects get counted as one source.

Such issues have generated considerable discussion in recent years. In
constructing a catalog of objects which appear in the HDF, we have tried to take a fairly
conservative approach, not digging deeply into the noise, and not
adopting complex algorithms for source classification, weighted
photometry, or merging and splitting of objects.  As the data are
public, we anticipate that others will generate their own catalogs.
Indeed, comparisons among the various catalogs should be 
instructive in revealing the strengths and deficiencies of the various
algorithms.

For the catalog we used a revised version of FOCAS \cite{JT81,Valdes82}. The program revisions are by \citeN{AS96},
and are intended to make the program more useful for HST and
ground-based infrared imaging.  A specific enhancement useful for our
application is the ability to adjust isophotal detection
thresholds as a function of position using variance maps.
This is important for mosaiced data like the HDF, where the
effective exposure time at the image boundaries is less than
that at field center due to the telescope dithering process.

We shall limit our discussion here only to the version
2 catalog, which supercedes the initial catalog released with
the HDF version 1 data.  Source detection and deblending was
carried out on the exposure--weighted sum of the F606W and F814W
version 2 drizzled images to provide maximum limiting depth.
The resulting catalog of objects is then applied to the registered
images in each of the four individual bandpasses, so that photometry
is carried out through identical apertures at all wavelengths.
The F606W+F814W summed image is significantly deeper than
any of the individual images, and few normal objects should be
missed by using it to define the object catalog for all bands.
It is possible, however, that objects with very strong emission
lines in the F450W or F300W bands could have escaped detection.

To identify sources, the data are first convolved with a fixed
smoothing kernel $K_{ij}$, shown in Table \ref{tabfocaskernel}.
Pixels with convolved values higher than a fixed threshold above
a local sky background are marked as potentially being part of
an object.  A single value for $\sigma_{\rm sky}$ was measured
for each CCD using a fit to the sky histogram, and then input as
a parameter to the FOCAS catalog for object detection.  FOCAS sets
the detection threshold $T$ to a constant times $\sigma_{\rm sky}$:
\begin{equation}
 T = N \sigma_{\rm sky} \left(\sum_{i,j} K_{i,j}^2\right)^{1\over 2}, 
\end{equation}
where the sum over the squares of the filter kernel elements
accounts for the noise supression {\it expected} for Poisson
sky noise from the smoothing process.  Note, however, that the
pixel--to--pixel noise in the drizzled images is correlated
(see \S 4.5), and therefore the {\it measured} pixel--to--pixel
{\it rms} provides an underestimate of the ``true'' noise level
of the image by a factor of 1.9.  Smoothing the drizzled data
does not suppress the noise as much as FOCAS expects for uncorrelated
noise.  We have therefore empirically adjusted the scaling factor
$N$ until the detection of spurious sources was minimized.
In the end, a threshold $N = 4$ was adopted.

Note also that the sky {\it rms} is (relatively) larger on
the PC than on the WFC detectors.  Therefore the isophotal threshold
in physical units (e.g. mag/square arcsecond) on the PC is somewhat
higher than that used for the WFC.
 
After thresholding, regions consisting of more than a certain number
of contiguous pixels (including diagonals) are counted as sources.
For the HDF data, the source detection threshould was set to $4\sigma$ and
the minimum area to 25 drizzled pixels,
or 0.04 square arcseconds.   This area is equivalent to 4
original WF pixels or 19.2 PC pixels.  It corresponds to an isophotal
diameter limit of roughly 0.2 arcsec, about 1.6 times larger than the
FWHM of the PSF on the WF cameras.
For a point source on the WFC, this minimum area encompasses
roughly 60\% of the total object flux.   Correspondingly, isophotal
fluxes of faint, extended galaxies with sizes close to the minimum
size limit must underestimate their total fluxes by at least 40\%.
 
These sources are then examined for sub-components using a variant
of the original detection scheme. The detection filter is repeatedly run
over the image as the threshold is raised. If, at some threshold, the
object breaks into two or more unconnected regions, each fragment which
meets the minimum area criterion of 25 pixels is given a
separate sub-entry in the catalog, and the process continues.
 
One additional test was applied to potential ``daughter'' objects during
the splitting process, using the FOCAS--defined ``significance'' parameter.
This parameter is a measure of surface brightness difference
between the object's peak and its detection isophote, in units of the
locally determined sky {\it rms} (see \cite{Valdes82} for details).
This ``significance test'' was not applied during the original detection
of parent objects, as it largely duplicates the original isophotal
thresholding criterion, but we have found empirically that its use helps
to suppress the over--splitting of spurious sources from the extended
wings of bright stars and large galaxies.  Therefore a significance
threshold of $> 5$ was imposed during object splitting.
 
The splitting process occasionally breaks large clumpy galaxies into
many individual sub-components, or produces many spurious sources along
the diffraction spikes of bright stars. We have fixed the most egregious
cases by manually merging the split objects back into their parent.
The cases for which this was done are flagged with an ``F'' in the catalog
below. In general, the only objects for which the splitting
was adjusted manually were bright, unmistakably recognizable spiral
galaxies, where FOCAS tends to sever off the arms and HII regions into
separate objects. These have been re--merged back into the parent.
There are many complex, irregular galaxies where we have allowed
the original FOCAS splitting to stand, despite the fact that in some
cases one's intuition might suggest that the divided object is in fact
a single entity.  As noted below in \S\ref{seccatalog}, the catalog presented here
includes {\it both parent and daughter objects.}  The user may
therefore adopt or reject the FOCAS splitting as he or she sees fit.
In section \ref{seccounts} below, we present one possible algorithm for ``re--merging''
over--split objects, and apply it to the issue of galaxy number counts.
 
\subsection{The Catalog}\label{seccatalog}
The source catalog is presented in Table \ref{tabcatalog}. 
For each object we report the following parameters:
\begin{description}
\item{\bf ID:\\} This is the FOCAS catalog entry number. The numbers after the
decimal point indicate the level of splitting. Both parents and 
daughters are included in the catalog shown here. Thus many objects
are included repeatedly in the catalog, both as part of the parent and
as a separate daughter entry.  For the statistical
distributions shown in the next section, we have adopted specific
color and separation criteria for merging objects to avoid double
counting.
\item{\bf x,y:\\} The x and y pixel positions of 
each object, as defined by brightest pixel within the $3 \times 3$ pixel grid
with the greatest luminosity within the detection area. For objects
with a bright off-center peak, this position can be significantly
different from the weighted center of the luminosity distribution within
the detection area. For the HDF, such differences are typically 
less than 0.1 arcsec.
\item{\bf RA,DEC:\\}
Minutes and seconds of the right ascension and declination 
corresponding to the x,y centers, epoch J2000. For RA these
are minutes and seconds of time. To these must be
added 12 hours (RA) and 62 degrees (Dec). 

The RA and Dec positions given in Table 7 were derived as follows.
The geometric distortion of each of the WFPC2 CCDs has been accurately 
measured (Trauger et al. 1995), and was removed in the drizzling stage of 
the image reduction process (see \S 4.5).  The relative positions of the
CCDs are somewhat less accurately determined, and indeed are known to 
drift with time.  WFPC2 image headers are encoded with a world coordinate
system tied to the HST guide star reference frame.  Absolute positions in
this reference frame may be incorrect by an arcsecond or more, depending on
the position on the sky, and on the accuracy of the individual positions for 
the guide stars used during the HST observations.

We have recalibrated the absolute zeropoint of the HDF coordinates using
interferometric radio positions for two sources within the HDF detected
by both the Merlin array and the VLA.   The radio astrometry for these 
was kindly provided by Tom Muxlow and Ken Kellerman, and the two independant 
radio measurements for each source agree with one another within their quoted 
uncertainties.  Here, we have adopted the Merlin positions, whose positional 
accuracy is of $\sim$20mas or better.  The WFPC2 coordinates for the optical 
counterparts of these two sources were compared to the Merlin positions, 
and a simple, mean translational offset ($\Delta\alpha = +0^s.089 \pm 0.010$ 
and $\Delta\delta = -1^{\prime\prime}.03$) was adopted and applied to the 
``raw'' WFPC2 coordinates. 
A global solution using VLA sources detected in both the central
field and the flanking fields \cite{Windhorst96}
yeilds a mean offset 0.4 arcseconds different from the one adopted
above, but consistent to within uncertainties of the measured source
coordinates.

The resulting coordinate system used for the HDF catalog should
therefore be accurate (within the radio reference frame provided by the
Merlin data) to approximately $0^{\prime\prime}.4$.  Relative positions
for individual galaxies are also typically accurate to
$0^{\prime\prime}.1$, with the largest uncertainty coming from the
ability of FOCAS to determine a peak position for faint, lumpy objects.


\item{\bf $m_t$,$m_i$:\\} The magnitudes of the detected
sources in the F606W image. These magnitudes are in the AB system, where
$m = -2.5 \log f_\nu -48.60$. \nocite{Oke74} 
The ``isophotal'' magnitude $m_i$ is
determined from the sum of the counts within the detection isophote. 
The ``total'' magnitude is computed from the number of counts
within a ``grown'' area. The total area is determined by first filling
in any x or y concavities in the isophote shape and then adding by rings around
the object until the area exceeds the detection area by at least a factor of
two. For daughter objects, the total magnitude is divided between the
daughters in proportion to their isophotal luminosities. The isophotal
magnitudes correspond to the higher isophotes at which the object broke into
multiple components. 

\item{\bf $u-b$, $b-v$, and $v-i$:\\}
Colors within the detection area. These are
essentially isophotal colors measured to a faint limiting isophote defined
from the summed F606W+F814W image. They are expressed in the AB system. 
(Our preferred notation for these colors is $U_{300} - B_{450}$, 
$B_{450} - V_{606}$, and $V_{606}-I_{814}$, to avoid confusion with 
the groundbased Johnson and Str\"omgren systems. However space in the
tables does not allow us to use this convention here.)
Galaxies where one band is a non-detection, as defined by
having signal-to-noise ratio $S/N < 2$ within one of the bands,
are marked as upper or lower limits (depending on which band drops out).
If both bands are upper limits, no color is given.

\item{\bf $S/N$:\\} The signal-to-noise ratio of the detection in the summed
F606W $+$ F814W image, based on
a semi-empirical noise model. If $L_i$ is the sky-subtracted number of counts
within the detection isophote, the expected variance is 
\begin{equation}
(\Gamma\sigma(L_i))^2 = \Gamma N_{\rm obj} + 1.9 \sigma_{\rm sky}^2 A_{\rm obj}
+ A_{\rm obj}^2 1.9 \sigma_{\rm sky}^2 / A_{\rm sky},
\end{equation}
where $\Gamma$ is the inverse gain (assumed to be 7 for all chips), 
$N_{\rm obj}$ is the total number of counts in the object aperture,
$A_{\rm obj}$ and $A_{\rm sky}$ are the number of pixels within the object 
and sky apertures, respectively,
and $\sigma_{\rm sky}$ is the measured standard deviation
of the level within the sky aperture. The first term accounts for 
Poisson variations in total counts in the source aperture. The second
term accounts for statistical variations in the mean sky level expected
from Poisson statistics, and the third term accounts for the random
(but not any systematic)
uncertainty in determining the mean sky level within the sky aperture.
The factor of 1.9 is an empirical correction to the measured
pixel-to-pixel standard deviation to account for the fact that the
images are in effect smoothed by the subsampling procedure. Therefore
the pixel-to-pixel variance underestimates the variance that will be
seen on large scales, as described above in \S\ref{secdrizzle}.  While
this factor is in principle a function of scale, our detection
apertures and sky apertures are large enough that a constant value is a
good approximation.
\item{\bf $A$:\\}
Area in pixels within the detection isophote.
\item{\bf $r_1$:\\}
Intensity-weighted first-moment radius determined from pixels within
the detection isophote. The radii are determined relative to the $x,y$
centers listed in the catalog, and the first moment radius is
\begin{equation}
r_1 = \sum r I(x,y) / \sum I(x,y), 
\end{equation}
where $I(x,y)$ is the intensity in each pixel.
\item{\bf PA:\\} 
The intensity-weighted position angle defined such that an object pointing
North-South has $\theta = 0$, and the position angle increases as the major axis
of the object rotates toward the east. In chip coordinates, the 
position angle is 
\begin{equation}
\theta_{xy} = 0.5 \tan^{-1}[2 Z/(Y-X)],
\end{equation}
where X, Y, and Z are defined from the second moments of the image, as follows:
\begin{equation}
X = \sum x^2 I(x,y) / \sum I(x,y), 
\end{equation}
\begin{equation}
Y = \sum y^2 I(x,y) / \sum I(x,y), \rm and
\end{equation}
\begin{equation}
Z = \sum xy I(x,y) / \sum I(x,y).
\end{equation}
The nominal spacecraft roll (PA\_V3) of $112^\circ$ was assumed in converting
from the measured position angle to a celestial position angle.
\item{\bf $b/a$:\\} 
The intensity-weighted axial ratio taken from the second moment 
of the light distribution. Define 
\begin{equation}
P = X + Y, \rm and
\end{equation}
\begin{equation}
Q = [(X-Y)^2 + 4 Z^2]^{1/2},
\end{equation}
where $X, Y$, and $Z$ are the moments defined above. Then the 
axial ratio is 
\begin{equation}
\epsilon = [(P-Q)/(P+Q)]^{1/2}.
\end{equation}
\item{\bf Flags:} 
S indicates that the source is a single object (not split into subcomponents).
B indicates that the outer isophote of the source overlaps a chip boundary in 
one or more bandpasses. F indicates that object components originally detected
by FOCAS were manually merged back into their parent as described above.
\end{description}

\section{Galaxy Counts}\label{seccounts}
While it is not our purpose to discuss scientific results in this
paper, it is useful to examine some of the basic statistics that will
be important for such analyses. 

The simplest statistic is the number of galaxies in the image. The counts
of galaxies as a function of apparent magnitude are an essential tool
of observational cosmology, and obtaining the faintest possible counts
was one of the prime motivations behind the HDF project. However, before
presenting the counts, it is useful to discuss the issue of object
splitting, which has been a source of some interest for the faintest
counts \cite{CROS96}.

Figure \ref{figfocasimage} shows a section of the F606W image on chip
4, with the FOCAS identifications labeled. Examples of objects for which
image splitting becomes problematical are sources 858, 774, and 555, which
are actually among the most interesting sources in the entire HDF.
Source 858 has been spectroscopically identified as a galaxy at
$z=3.226$ \cite{SGDA96}.  The spectrum refers to all four components,
which have essentially the same color in the HDF image. Source 774
consists of three components which all have the same, very blue color.
Source 555 consists of multiple components with different colors. The
compact object 555.2 is very red and is probably a moderate-redshift
elliptical galaxy. The elongated structure is spectroscopically
identified as a galaxy at $z = 2.803$ \cite{SGDA96}. FOCAS splits it
into two objects 555.11 and 555.12. These subcomponents have nearly
identical colors. In the catalog, we list the parents and all the
daughters for these objects. {\it Many objects are therefore
double-counted in the catalog.} In counting the galaxies, we must
decide which components of the merged objects to keep separate, and
which to count as a single object. The separations of the components of
858 and 747, for example, are under 100 kpc, even for an open
cosmology. They are thus probably more fairly considered as pieces of a
larger galaxy, rather than as separate entities. However, a simple
separation criterion will often merge objects such as 555.1 and 555.2
that are likely to be chance projections rather than physical associations. We have
therefore adopted criteria for merging daughters that combine both color
and separation. Specifically, we require that 
$\Delta (V_{606} - I_{814}) < C$ and that the separation $S_{ij}$ between two
components be
\begin{equation}
        S_{ij} < F \times (r_i+r_j), 
\end{equation}
where the radii are determined from the FOCAS isophotal area as
$r_i = \sqrt{A/\pi}$ and the factor $F$ is a tuneable parameter.
The values $C = 0.3$ and $F=5$ were chosen after some experimentation.
These values merge all the components of sources 858 and 774, and
merge the blue components of 555, leaving the red component as 
a separate object. 

It is clearly a matter of subjective judgement whether to merge such objects.
Does it matter? Table 
\ref{tabsplitting} compares the counts in three magnitude intervals
for our chosen values $F=5, C=0.3$; for 
values $F=0, C=0$, which provide maximal splitting (keeping only the 
daughters); and for
$F=100, C=10$, which provide maximal merging (keeping only the parents).
The change in counts from maximal to minimal merging is less than 20\% 
in any magnitude interval. Thus splitting
issues, while interesting and important for studies assessing the 
sizes, clustering, and morpohology of faint galaxies, have only a small
influence on the overall counts.

Incompleteness and uncertainties in the magnitude estimates also influence
the counts. These are best addressed by means of simulations, and will
be discussed in detail in \citeN{Ferg96marginal}. For the
purposes of this paper, it should be noted that FOCAS isophotal, total,
and aperture magnitudes are each subject to certain systematic biases.
The isophotal magnitudes, in particular, are likely to be approximately
0.2 mag fainter than total magnitudes for $V_{606} > 28$.
The bias for total magnitudes is smaller but still
non-negligible. The counts presented here have not been corrected
for these biases. The counts are presented down to magnitudes at which
they are likely to be more than 80\% complete (under certain 
standard assumptions for the intrinsic surface-brightness distribution 
-- see \citeNP{Ferg96marginal} for details). 

Figures \ref{fig300counts} - \ref{fig814counts} show the HDF counts
in the individual bands.  Isophotal, total, and aperture
magnitudes are shown separately to illustrate the effects of different
magnitude estimates.  The counts in isophotal and total magnitudes
are tabulated in Tables \ref{tabcountsub} and \ref{tabcountsvi}.  
Figure \ref{fignmBI} shows the F450W and F814W band counts together with
ground-based data. All the surveys have been corrected to AB
magnitudes, but no color corrections have been applied.

The essential shape of the counts is relatively robust.  In all bands
the slopes flatten at faint magnitudes.  The slopes in several
magnitude intervals are given in Table \ref{tabslopes} (computed from simple
least-squares fitting of the binned data over the magnitude intervals
shown). The rather sharp flattening in the F300W counts appears to be
unlikely to be due to incompleteness \cite{Ferg96marginal}, and quite
possibly indicates that many of the galaxies at faint magnitudes have
redshifts $z > 2$ such that they drop out of the F300W band due to the
combined affects of internal extinction (which can be quite severe in
the rest-frame far-UV), internal Lyman continuum absorption, and
intergalactic absorption from the Lyman $\alpha$ forest and from
Lyman-limit systems \cite{SGDA96,MFDGSF96}.  The flattening of
the counts in the F450W band may signal the loss of galaxies at $z > 3$
due to the same effects.

Figures \ref{figcmUB} -- \ref{figcmVI} show color--magnitude diagrams
for galaxies in the field. For comparison, also shown are the 
colors for template non-evolving galaxies of different types.
The templates for E, Sbc, and Im galaxies are taken from \citeN{CWW80},
with the extrapolations adopted by \citeN{FM95}. We have normalized
the E and Sbc spectra to have absolute magnitudes in the F450W band of
$B_{450} = -21.1$, roughly $L^*$ for $H_0 = 50 \rm \,km\,s^{-1} Mpc^{-1}$.
The Im spectrum is normalized to an absolute magnitude $B_{450} = -18$,
more typical of a Magellanic irregular. The curves are labeled with
redshift for galaxies of these assumed absolute magnitudes. 
It is interesting to note that while the elliptical galaxy template 
outlines the red envelope of the data reasonably well, there is a 
large population of galaxies at faint magnitudes that are bluer than
the irregular galaxy template at any redshift. Also, as noted for
the counts, there is a substantial number of galaxies with
$U_{300} - B_{450}$ and $B_{450} - V_{606}$ colors and magnitudes
consistent with being luminous $(\gtrsim L^*)$ galaxies at high
redshift $(z > 2)$.

\section{Conclusions}
The Hubble Deep Field observations were taken with the expectation 
that they
will contribute to the resolution of some of the outstanding
questions in studies of galaxy formation. This paper 
describes the motivation, field and filter selection, and
data reduction. Data, catalogs, and
further information on the project are available on the 
World-Wide Web at http://www.stsci.edu.

\acknowledgments
We are extremely grateful to the large number of people who have
contributed to the HDF project. In particular, we would like
to acknowledge the efforts of the Advisory Committee, who 
donated a significant amount of effort in advising how best to
use the Director's discretionary time, and in offering suggestions
on how to carry out the observations. Among the many members of the STScI
staff who helped with the observations, we would like to acknowledge 
especially J. C. Hsu and Ivo Busko for their timely 
efforts in devoloping software for data reduction. We thank Rogier
Windhorst and Tom Muxlow for providing details on the comparison of 
radio and optical coordinates within the HDF.


\begin{deluxetable}{ll}
\tablecaption{\label{tabfield}
Characteristics of the Hubble Deep Field
}
\startdata
Location: & 12h 36m 49.4s  $+62^\circ 12' 58"$  \nl
 & (Epoch J2000.0 / WFPC-2 'WFALL FIX' position) \nl
 & V3 Position angle = $112^\circ$ \nl
$E(B-V)$: & 0.00 \nl
HI column density: & $ \rm 1.7 \times 10^{20}  cm^{-2}$ \nl
DIRBE flux: &  $\rm < 0.14 MJy/ster$ \nl
Radio sources: & none with flux $>$ 1 mJy at 3.6 cm \nl
IRAS cirrus: & Local minimum in 100$\mu \rm m$ maps \nl
Bright stars: & None near the field \nl
Galaxy Clusters: & Nearest is 48 armin away \nl
\enddata
\end{deluxetable}

\begin{deluxetable}{lrrrl}
\tablecaption{\label{tabflanking}
HDF Flanking Fields}
\tablehead{
\colhead{Equatorial}               & \colhead{}      &
\colhead{}       & \colhead{Total}  &
\colhead{}  \\
\colhead{J2000 Coordinates}               & \colhead{UT Date}      &
\colhead{N$_{\rm exp}$}       & \colhead{T$_{\rm exp}$ (s)}  &
\colhead{Comments} 
}
\startdata
12 36 35.17 +62 13 38.7 & 30 Dec 1995 & 4 & 5300s & Inner West \nl
12 36 20.93 +62 14 19.3 & 18 Dec 1995 & 3 & 2500s & Outer West \nl
12 37 03.62 +62 12 17.2 & 30 Dec 1995 & 4 & 5300s & Inner East \nl
12 37 17.83 +62 11 36.3 & 29 Dec 1995 & 3 & 3000s & Outer East \nl
12 36 51.03 +62 15 47.6 & 30 Dec 1995 & 3 & 2500s & North West \nl
12 37 05.27 +62 15 06.8 & 29 Dec 1995 & 3 & 2500s & North East \nl
12 36 47.77 +62 10 08.4 & 29 Dec 1995 & 3 & 2500s & South East \nl
12 36 33.56 +62 10 49.1 & 30 Dec 1995 & 3 & 2500s & South West \nl
\enddata
\end{deluxetable}

\begin{deluxetable}{lrrr}
\tablewidth{3.0in}
\tablecaption{\label{tabmatrix}
Sky Noise Correlation Matrix
}
\tablehead{ \colhead{} & \colhead{$i-1$} & \colhead{$i$} & \colhead{$i+1$} }
\startdata
$j-1$ & 0.11 & 0.33 & 0.11 \nl
$j$ & 0.33 & 1.00 & 0.33 \nl
$j+1$ & 0.11 & 0.33 & 0.11 \nl
\enddata
\end{deluxetable}

\begin{deluxetable}{lrrr}
\tablewidth{3.0in}
\tablecaption{\label{tabnoisekernel}
Noise Kernel}
\tablehead{ \colhead{} & \colhead{$i-1$} & \colhead{$i$} & \colhead{$i+$1} }
\startdata
$j-1$ & 0.06 & 0.18 & 0.06 \nl
$j$ & 0.18 & 0.93 & 0.18 \nl
$j+1$ & 0.06 & 0.18 & 0.06 \nl
\enddata
\end{deluxetable}

\begin{deluxetable}{rrrcrrrr}
\small
\tablecaption{\label{tabexptimes} HDF Image Parameters}
\tablehead{
\colhead{} & \colhead{Number of} & \colhead{Exposure} & \colhead{10$\sigma$ AB mag} &
\colhead{Sky\tablenotemark{a}} & \colhead{Sky} & \colhead{Sky} & \colhead{Sky} \\
\colhead{Filter} & \colhead{Frames} & \colhead{Time (s)} & \colhead{limit} &
\colhead{PC1} & \colhead{WF2} & \colhead{WF3} & \colhead{WF4}
}
\startdata
   F300W  &  77  &  153700 & 26.98 &  39.259 &   78.642 &   92.954 &   89.686 \nl
   F450W  &  58  &  120600 & 27.86 &  65.055 &  310.133 &  319.222 &  309.483 \nl
   F606W  & 103  &  109050 & 28.21 & 238.507 & 1146.402 & 1155.758 & 1131.676 \nl
   F814W  &  58  &  123600 & 27.60 & 158.615 &  768.667 &  771.862 &  754.104 \nl
\enddata
\tablenotetext{a}{Sky values are total DN for the exposure times listed.}
\end{deluxetable}

\begin{deluxetable}{lccccc}
\tablewidth{3.0in}
\tablecaption{\label{tabfocaskernel}
FOCAS Smoothing Kernel}
\tablehead{ \colhead{} & \colhead{$i-2$} & \colhead{$i-1$} & 
\colhead{$i$} & \colhead{$i+1$} & \colhead{$i+2$} }
\startdata 
$j-2$ & 0 & 1 & 2 & 1 & 0 \nl
$j-1$ & 1 & 3 & 3 & 3 & 1 \nl
$j$ & 2 & 3 & 4 & 3 & 2 \nl
$j+1$ & 1 & 3 & 3 & 3 & 1 \nl
$j+2$ & 0 & 1 & 2 & 1 & 0 \nl
\enddata
\end{deluxetable}

\begin{deluxetable}{l}
\tablecaption{\label{tabcatalog}
HDF Catalog
}
\tablehead{ \colhead{} }
\startdata
See attached pages. 
\enddata
\end{deluxetable}

\begin{deluxetable}{lrrr}
\tablecaption{\label{tabsplitting}
Counts for Various Amounts of Splitting}
\tablehead{ \colhead{$V_{606}$} & \colhead{$F=5$} & \colhead{$F=0$} & \colhead{$F=100$} \\
\colhead{range} & \colhead{$C=0.3$} & \colhead{$C=0$} & \colhead{$C=10$} }
\startdata 
$23 - 25$ & 137 & 128 & 133 \nl
$25 - 27$ & 491 & 528 & 450 \nl
$27 - 29$ & 1131 & 1239 & 1050 \nl
\enddata
\end{deluxetable}

\begin{deluxetable}{rrrrrrr}
\tablecaption{\label{tabcountsub}
HDF Galaxy Counts --- F300W and F450W bands}
\tablehead{ 
\colhead{} & 
\multicolumn{3}{c}{$U_{300}$} & \multicolumn{3}{c}{$B_{450}$} \\
\cline{2-4} \cline{5-7} \\ 
\colhead{} & \colhead{} & \colhead{isophotal} & \colhead{total} &
\colhead{} & \colhead{isophotal} & \colhead{total} \\
\colhead{AB} & \colhead{} & \colhead{$\log(n)$} & \colhead{$\log(n)$} &
\colhead{} & \colhead{$\log(n)$} & \colhead{$\log(n)$} \\
\colhead{mag} & 
\colhead{N} & \colhead{$\rm mag^{-1}\, deg^{-2}$} & \colhead{$\rm mag^{-1}\, deg^{-2}$} &
\colhead{N} & \colhead{$\rm mag^{-1}\, deg^{-2}$}  & \colhead{$\rm mag^{-1}\, deg^{-2}$} 
}
\startdata 
   22.25 &      3 &     3.60 &     3.42 &      7 &     3.97 &     4.03 \nl
   22.75 &      4 &     3.73 &     3.97 &      6 &     3.90 &     3.97 \nl
   23.25 &      6 &     3.90 &     3.60 &      9 &     4.08 &     4.03 \nl
   23.75 &     11 &     4.17 &     4.17 &     17 &     4.35 &     4.38 \nl
   24.25 &     15 &     4.30 &     4.33 &     28 &     4.57 &     4.66 \nl
   24.75 &     16 &     4.33 &     4.45 &     42 &     4.75 &     4.72 \nl
   25.25 &     40 &     4.72 &     4.83 &     69 &     4.96 &     5.03 \nl
   25.75 &     57 &     4.88 &     4.82 &     75 &     5.00 &     4.99 \nl
   26.25 &     73 &     4.99 &     5.04 &    107 &     5.15 &     5.22 \nl
   26.75 &     77 &     5.01 &     5.03 &    135 &     5.25 &     5.27 \nl
   27.25 &     89 &     5.07 &     5.11 &    150 &     5.30 &     5.32 \nl
   27.75 &     78 &     5.01 &     4.92 &    167 &     5.35 &     5.43 \nl
   28.25 & \nodata & \nodata & \nodata &    246 &     5.51 &     5.56 \nl
   28.75 & \nodata & \nodata & \nodata &    248 &     5.52 &     5.54 \nl
   29.25 & \nodata & \nodata & \nodata & \nodata & \nodata & \nodata \nl
\enddata
\end{deluxetable}

\begin{deluxetable}{rrrrrrr}
\tablecaption{\label{tabcountsvi}
HDF Galaxy Counts --- F606W and F814W bands}
\tablehead{ 
\colhead{} & 
\multicolumn{3}{c}{$V_{606}$} & \multicolumn{3}{c}{$I_{814}$} \\
\cline{2-4} \cline{5-7} \\ 
\colhead{} & \colhead{} & \colhead{isophotal} & \colhead{total} &
\colhead{} & \colhead{isophotal} & \colhead{total} \\
\colhead{AB} & \colhead{} & \colhead{$\log(n)$} & \colhead{$\log(n)$} &
\colhead{} & \colhead{$\log(n)$} & \colhead{$\log(n)$} \\
\colhead{mag} & 
\colhead{N} & \colhead{$\rm mag^{-1}\, deg^{-2}$} & \colhead{$\rm mag^{-1}\, deg^{-2}$} &
\colhead{N} & \colhead{$\rm mag^{-1}\, deg^{-2}$}  & \colhead{$\rm mag^{-1}\, deg^{-2}$} 
}
\startdata 
   22.25 &      7 &     3.97 &     3.97 &     14 &     4.27 &     4.20 \nl
   22.75 &      8 &     4.03 &     4.12 &     18 &     4.38 &     4.42 \nl
   23.25 &     15 &     4.30 &     4.33 &     28 &     4.57 &     4.60 \nl
   23.75 &     28 &     4.57 &     4.64 &     27 &     4.55 &     4.64 \nl
   24.25 &     43 &     4.76 &     4.72 &     48 &     4.80 &     4.85 \nl
   24.75 &     41 &     4.74 &     4.78 &     70 &     4.97 &     4.96 \nl
   25.25 &     82 &     5.04 &     5.08 &     84 &     5.05 &     5.13 \nl
   25.75 &     90 &     5.08 &     5.10 &    118 &     5.20 &     5.19 \nl
   26.25 &    139 &     5.27 &     5.26 &    134 &     5.25 &     5.28 \nl
   26.75 &    145 &     5.29 &     5.33 &    174 &     5.36 &     5.40 \nl
   27.25 &    191 &     5.40 &     5.43 &    196 &     5.42 &     5.43 \nl
   27.75 &    193 &     5.41 &     5.47 &    232 &     5.49 &     5.60 \nl
   28.25 &    330 &     5.64 &     5.68 &    343 &     5.66 &     5.71 \nl
   28.75 &    344 &     5.66 &     5.76 &    391 &     5.71 &     5.78 \nl
   29.25 &    439 &     5.77 &     5.83 & \nodata & \nodata & \nodata \nl
 
\enddata
\end{deluxetable}

\begin{deluxetable}{lcccc}
\tablecaption{\label{tabslopes}
Counts Slope}
\tablehead{ 
\colhead{mag range} &
\colhead{$U_{300}$} & \colhead{$B_{450}$} &
\colhead{$V_{606}$} & \colhead{$I_{814}$}
}
\startdata 
$23 - 26$ & 0.40 & 0.39 & 0.35 & 0.31 \nl
$26 - 29$ & 0.05\tablenotemark{a} & 0.16 & 0.17 & 0.18 \nl
\enddata
\tablenotetext{a}{The F300W limit here is 26-28; it is 26-29 for the other filters.}
\end{deluxetable}

\clearpage
\bibliography{ajmnemonic,bib} 
\bibliographystyle{aj}


\clearpage

\begin{figure}

\caption{\label{figfield}
The HDF and flanking fields, superimposed on a ground-based 
300 s $R$-band image taken with the Mayall 4-m telescope.
The small labeled insets show centers of each flanking field
image. The naming convention for the fields is given in
the last column of table 2.
}

\caption{\label{figfilters}
The bandpasses of the four filters chosen for the HDF. The total
system throughput is shown, including the contribution from the
filter, telescope and camera optics, and the detector.
}

\caption{\label{figbackground}
Predicted background as a function of time for representative
orbits (indicated as year.day)
during early, middle, and late parts of the HDF observing period. 
The background model 
includes scattered light from the illuminated earth on the 
day side of the orbit, which causes the large modulation shown
in the figure. The floor at $5 \times 10^{-7}$ is from the zodiacal
background and the excursion to zero is during earth-occultation.
}


\caption{\label{figdither}
HDF pointing positions.
Nine dither positions were planned to 
cover the large scale (left panel) offsets simultaneously and to provide good
sub-pixel sampling (right panel).  The offsets are
shown for the center of WF2 (after rotating x, y by 180 degrees
to match WF4).  The projection to WF (100 mas) sub-pixel
locations varies significantly from
chip to chip due to minor rotations and plate scale differences, 
as well as with position within a CCD due to differential
geometric distortion.  The symbols ``*", ``o", and ``+" refer to
the first, middle and last third of the exposure set for the 
F606W filter set plotted.  A drift of about 10 mas occurred over
multiple days.  As noted in the text, dither positions 10 and
11 resulted from anomalous Fine Guidance Sensor tracking, and
these frames also have a 4.3 arcminute rotation with respect to all the 
other data.
}

\caption{\label{figfootprint}
This figure illustrates the ``footprint'' of a pixel in the drizzling
process. The dark
region in the central pixel of this $3 \times 3$ grid shows the
smaller area adopted for the pixel when it is projected onto the
subsampled image.
}
\end{figure}
\begin{figure}

\caption{\label{figdrizzle}
This figure illustrates the major operations of drizzling. The
location of each input pixel in the output subsampled image is
determined from knowledge of the pointing position, rotation, and
geometric distortion. The flux within that pixel is then ``drizzled''
into the overlapping output pixels in quantities proportional to
the area of overlap.
}

\caption{\label{figpsf}
The point-spread function of a bright star in the final drizzled
HDF image.
}

\caption{\label{figcolor} A color composite image of the full HDF
field, constructed from the F450W,  F606W, and F814W images.}

\caption{\label{figF300_1} PC image in the F300W band. This and the following
images are from the version 2 drizzled images. Exposure times are given 
in Table 5. The pixel scale at the borders
is provided to allow galaxies in the catalog to be located via their x,y positions.}
\caption{\label{figF450_1} PC image in the F450W band.  }
\caption{\label{figF606_1} PC image in the F606W band.  }
\caption{\label{figF814_1} PC image in the F814W band.  }
\caption{\label{figF300_2} WF2 image in the F300W band.  }
\caption{\label{figF450_2} WF2 image in the F450W band.  }
\caption{\label{figF606_2} WF2 image in the F606W band.  }
\caption{\label{figF814_2} WF2 image in the F814W band.  }
\caption{\label{figF300_3} WF3 image in the F300W band.  }
\caption{\label{figF450_3} WF3 image in the F450W band.  }
\caption{\label{figF606_3} WF3 image in the F606W band.  }
\caption{\label{figF814_3} WF3 image in the F814W band.  }
\end{figure}
\begin{figure}
\caption{\label{figF300_4} WF4 image in the F300W band.  }
\caption{\label{figF450_4} WF4 image in the F450W band.  }
\caption{\label{figF606_4} WF4 image in the F606W band.  }
\caption{\label{figF814_4} WF4 image in the F814W band.  }

\caption{\label{figfocasimage} 
A section of the F606W image from WF4, with the FOCAS identifications marked.
}

\caption{\label{fig300counts}
Galaxy counts as a function of AB magnitude in the F300W band. 
FOCAS total and isophotal magnitudes are shown for $22 < U_{300} < 28$.
Aperture magnitudes are shown only for galaxies fainter than
$U_{300} = 26$, because the brighter galaxies are almost all larger
than the 0.5 arcsec radius aperture.
}

\caption{\label{fig450counts}
Galaxy counts as a function of AB magnitude in the F450W band. 
FOCAS total and isophotal magnitudes are shown for $22 < B_{450} < 29$.
Aperture magnitudes are shown only for galaxies fainter than
$B_{450} = 26.$
}

\caption{\label{fig606counts}
Galaxy counts as a function of AB magnitude in the F606W band. 
FOCAS total and isophotal magnitudes are shown for $22 < V_{606} < 29.5$.
Aperture magnitudes are shown only for galaxies fainter than
$V_{606} = 26.$
}

\caption{\label{fig814counts}
Galaxy counts as a function of AB magnitude in the F814W band. 
FOCAS total and isophotal magnitudes are shown for $22 < I_{814} < 29$.
Aperture magnitudes are shown only for galaxies fainter than
$I_{814} = 26.$
}

\end{figure}
\begin{figure}

\caption{\label{fignmBI}
Galaxy counts as a function of AB magnitude in the F450W and F814W bands,
together with a compilation of existing ground-based data. 
FOCAS total and isophotal magnitudes are shown for $22 < I_{814} < 29.$
No color corrections have been applied to the ground-based data.
}

\caption{\label{figcmUB}
Color-magnitude diagram $U_{300} - B_{450}$. Galaxies detected at
more than $5 \sigma$ are shown as large hexagons. Galaxies 
detected at less than $5 \sigma$ are smaller hexagons. Galaxies
either undetected in F300W or detected at less than $2\sigma$ 
significance are shown as open triangles at the position of
the $2 \sigma$ limit on the color. Also shown are the colors of
fiducial non-evolving spectra of E,
Sbc, and Im galaxies (see text), as dotted, solid, and dashed
curves, respectively.
}

\caption{\label{figcmBV}
Color-magnitude diagram $B_{450} - V_{606}$. The meanings of
the lines and symbols are the same as for the previous figure.
Triangles are shown at the position of the F450W $2\sigma$ limits.
}
\caption{\label{figcmVI}
Color-magnitude diagram $V_{606} - I_{814}$. The meanings of the
lines and symbols are the same as for the previous two figures. 
Triangles are shown at the position of the F606W $2\sigma$ limits.
}
\end{figure}
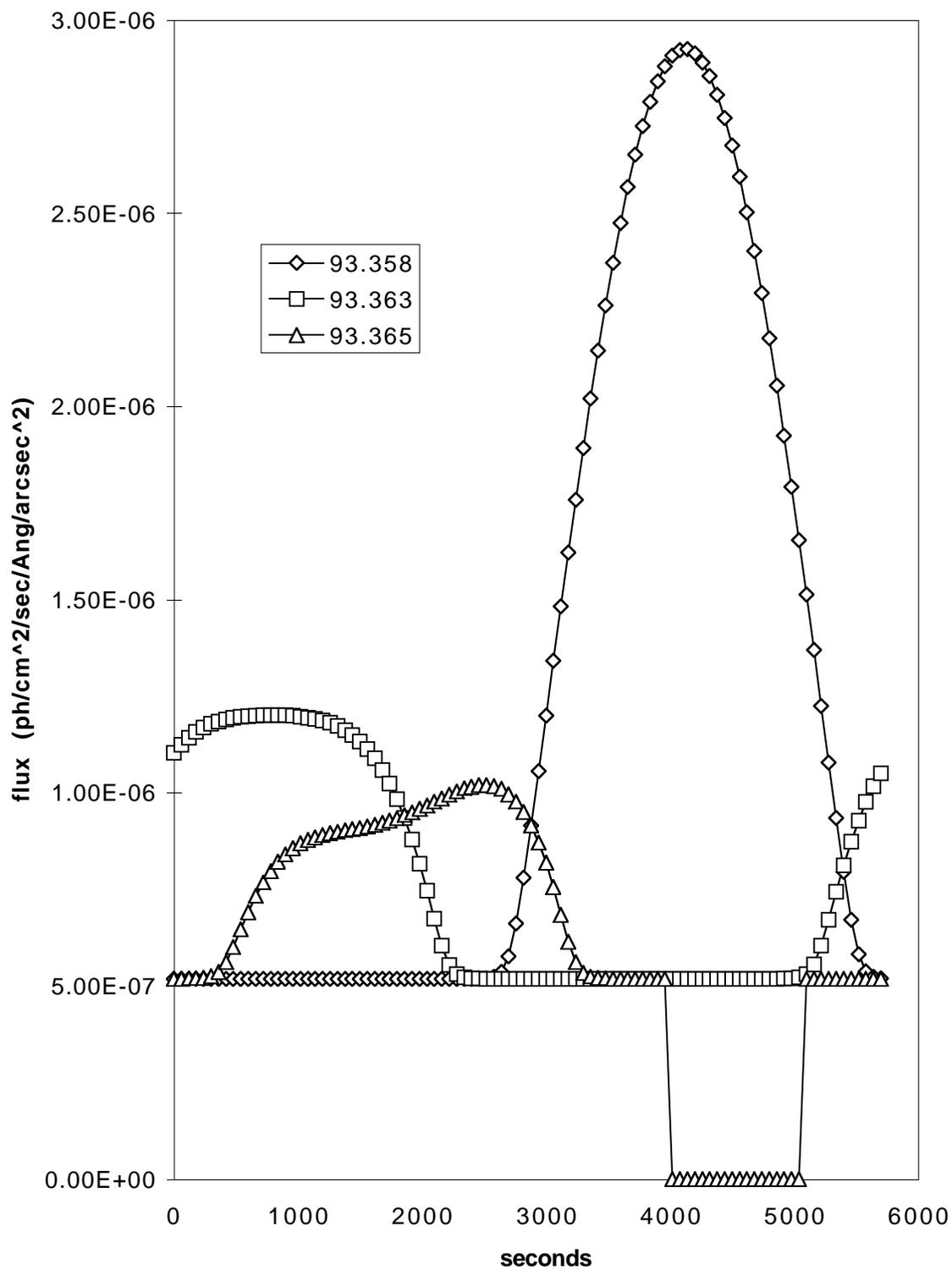
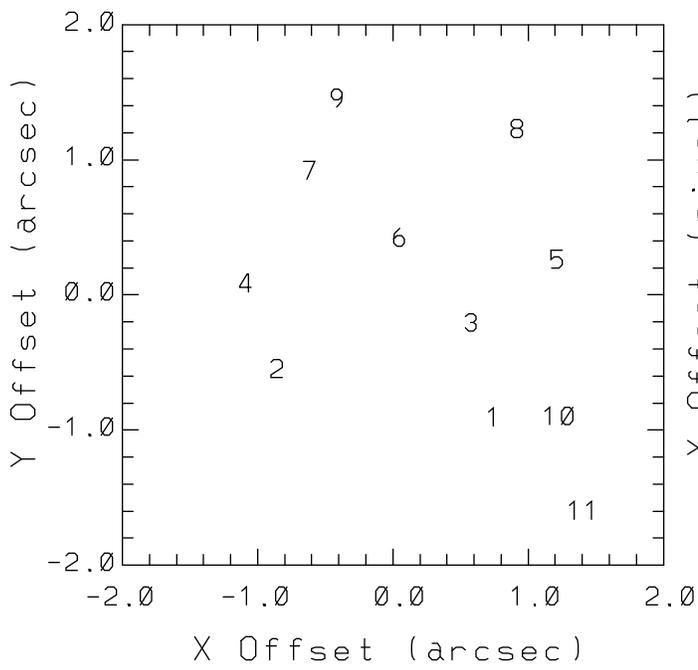
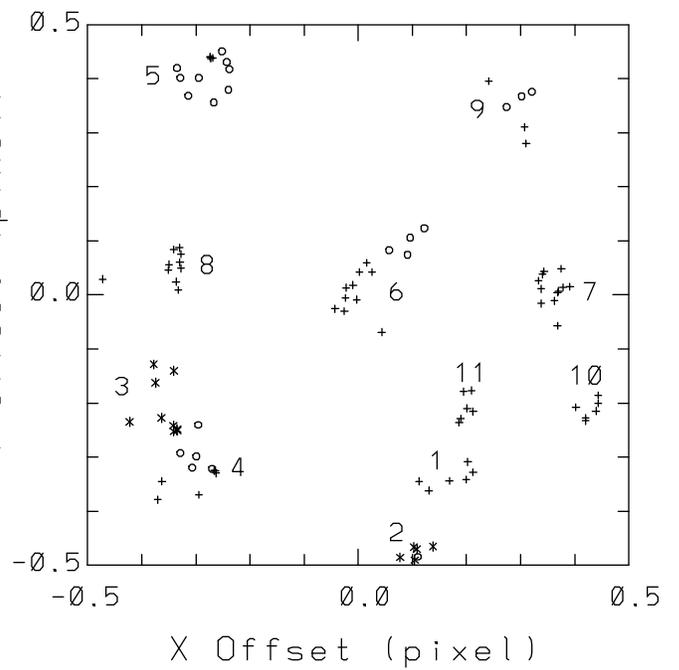
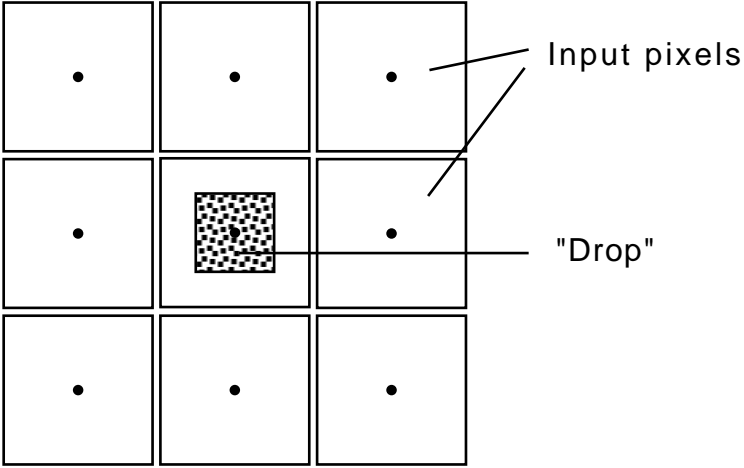
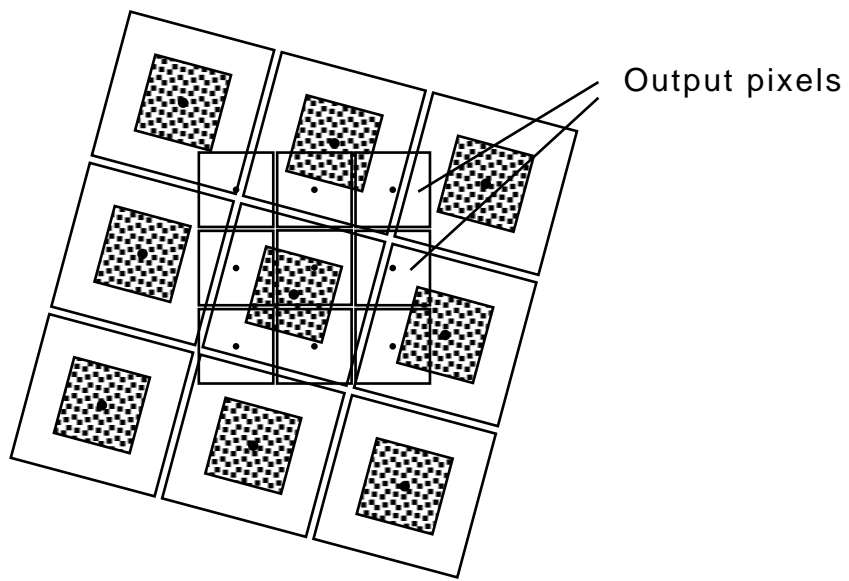
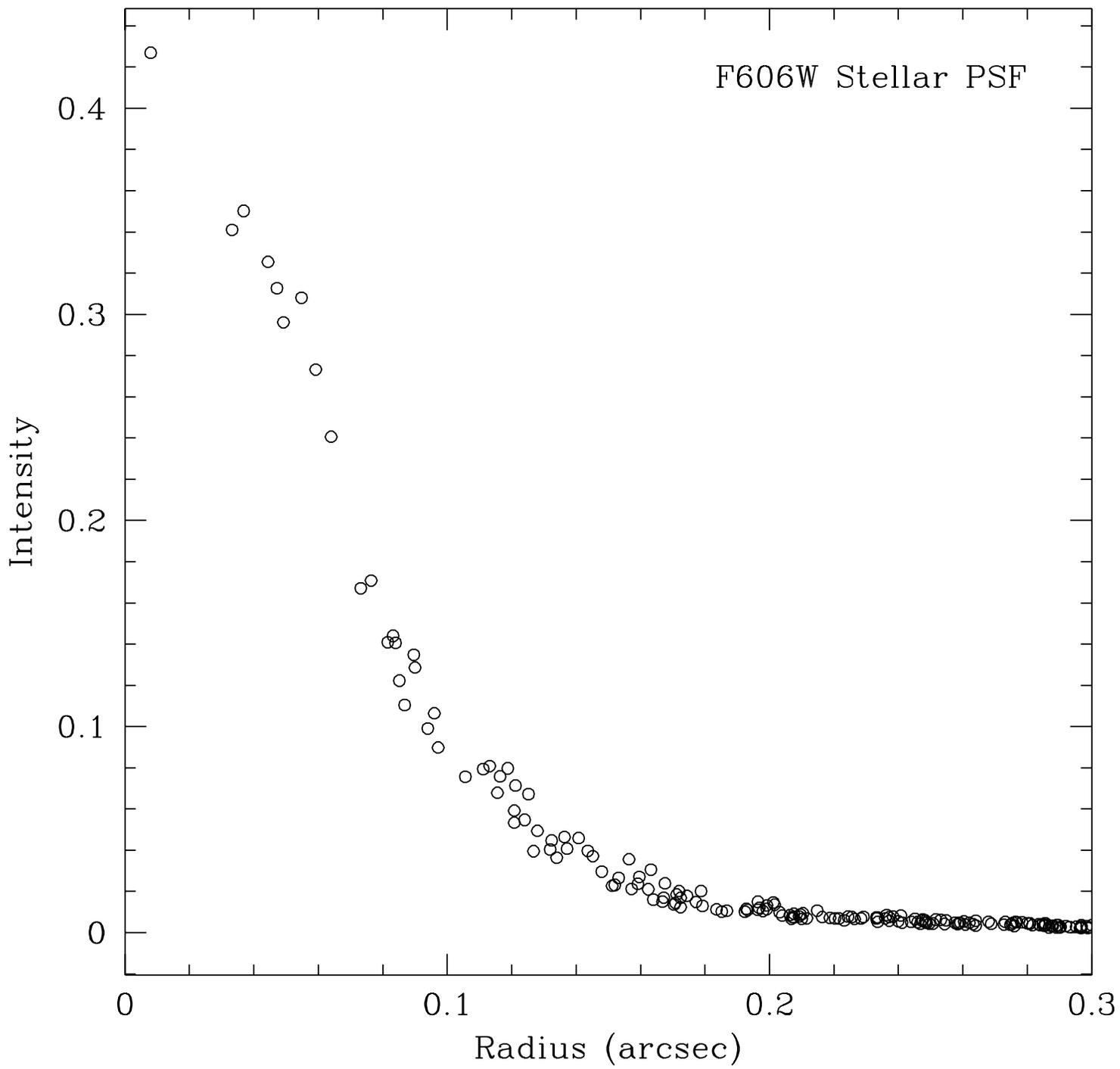
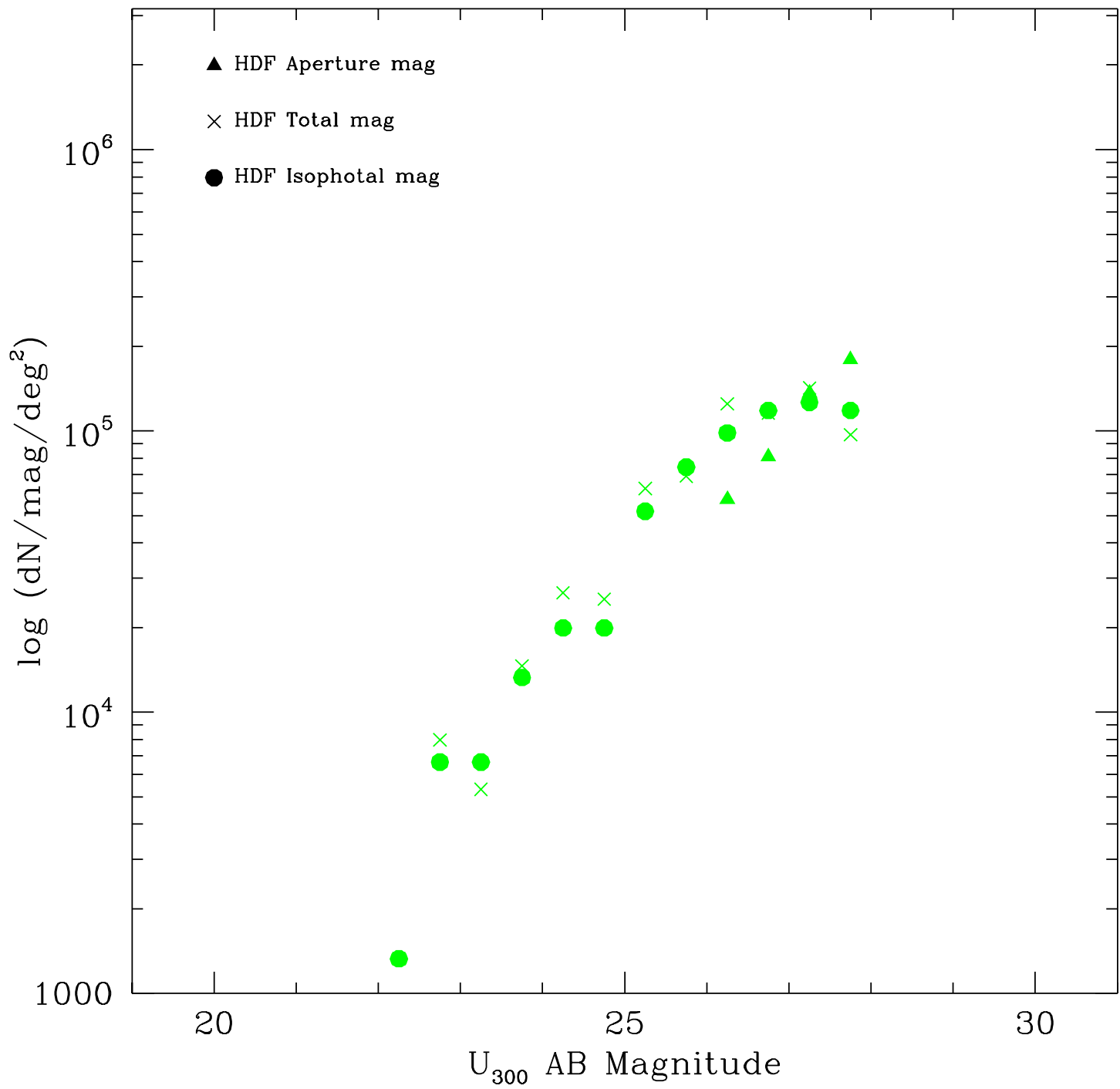
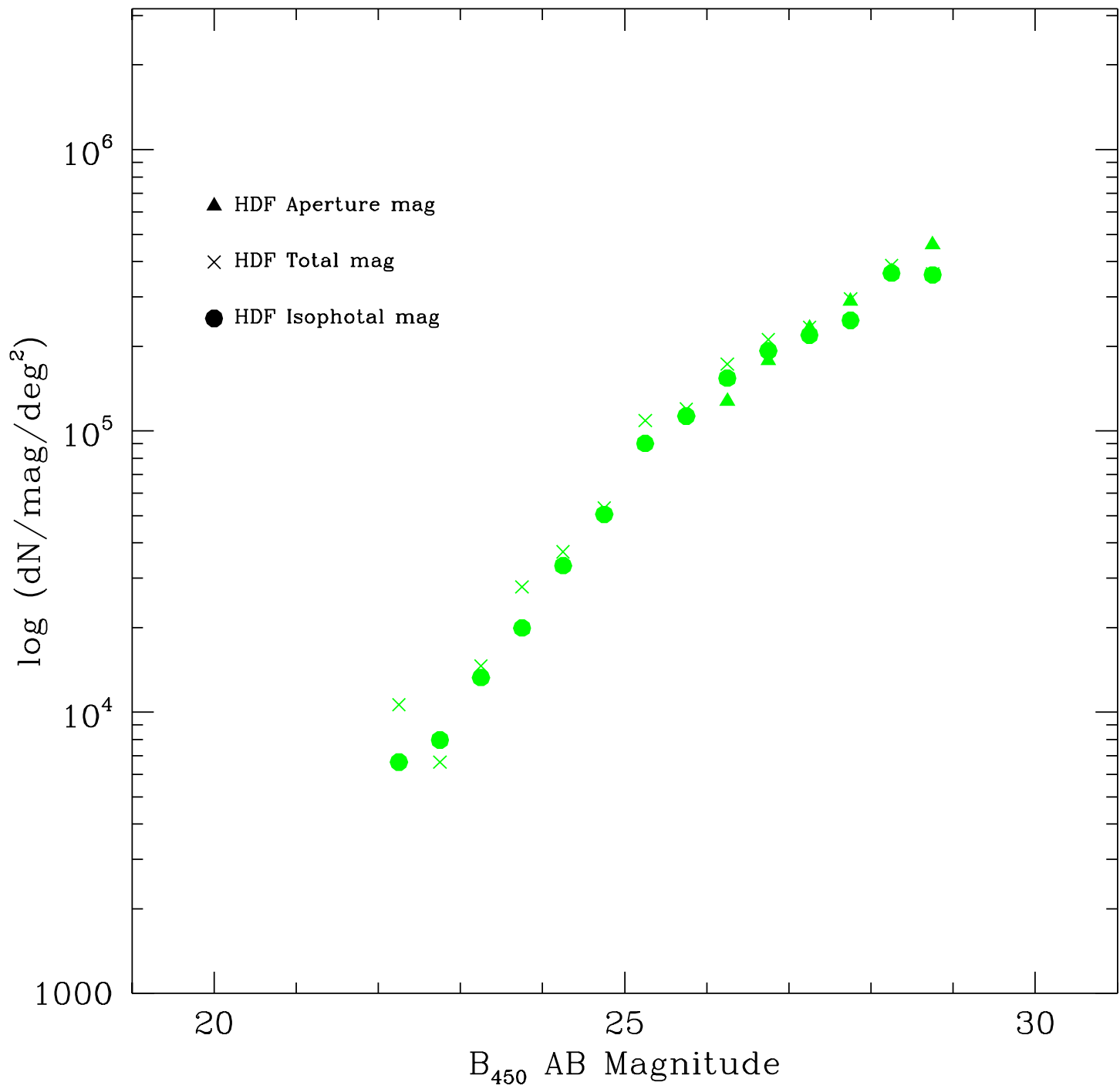
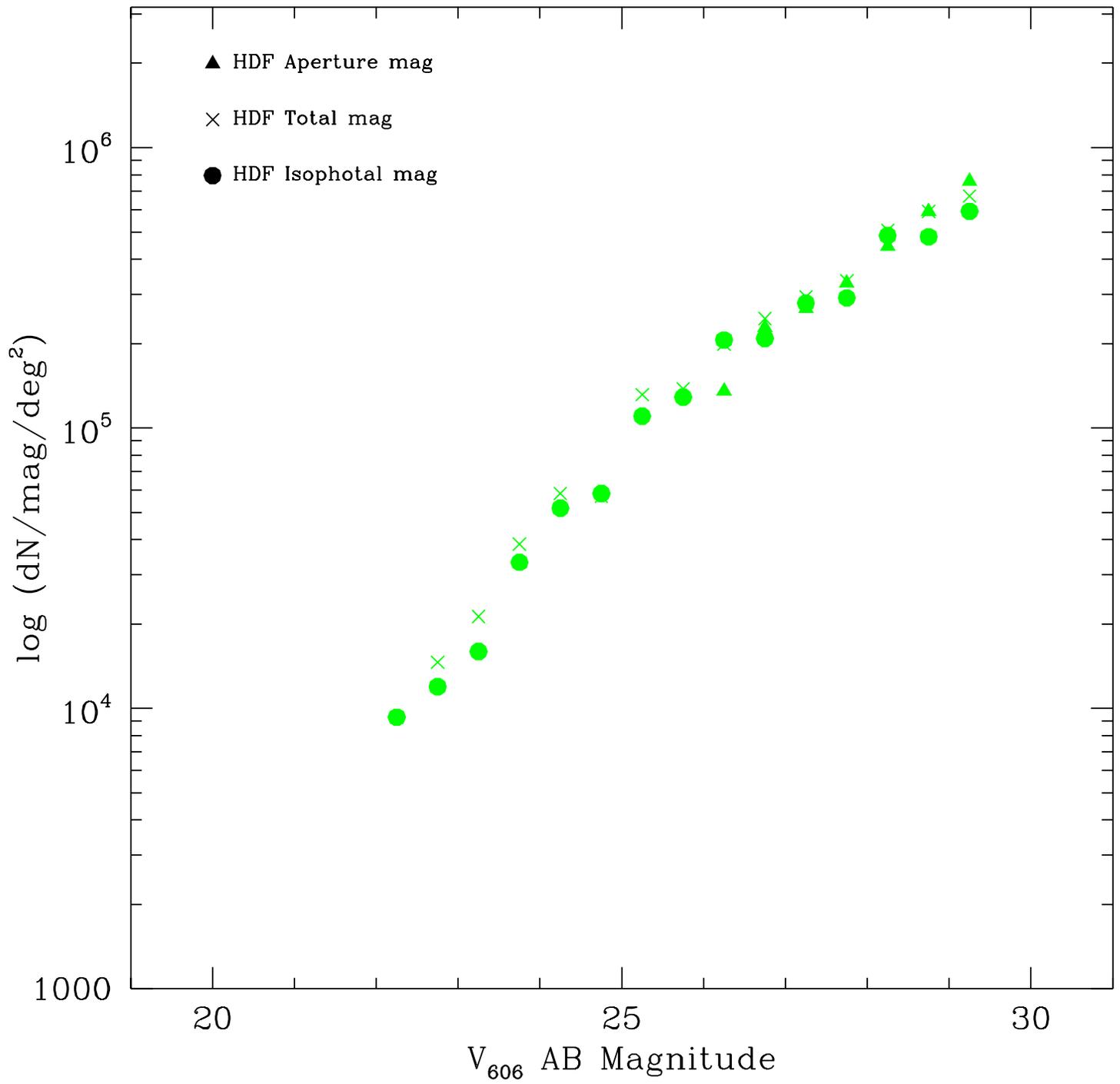
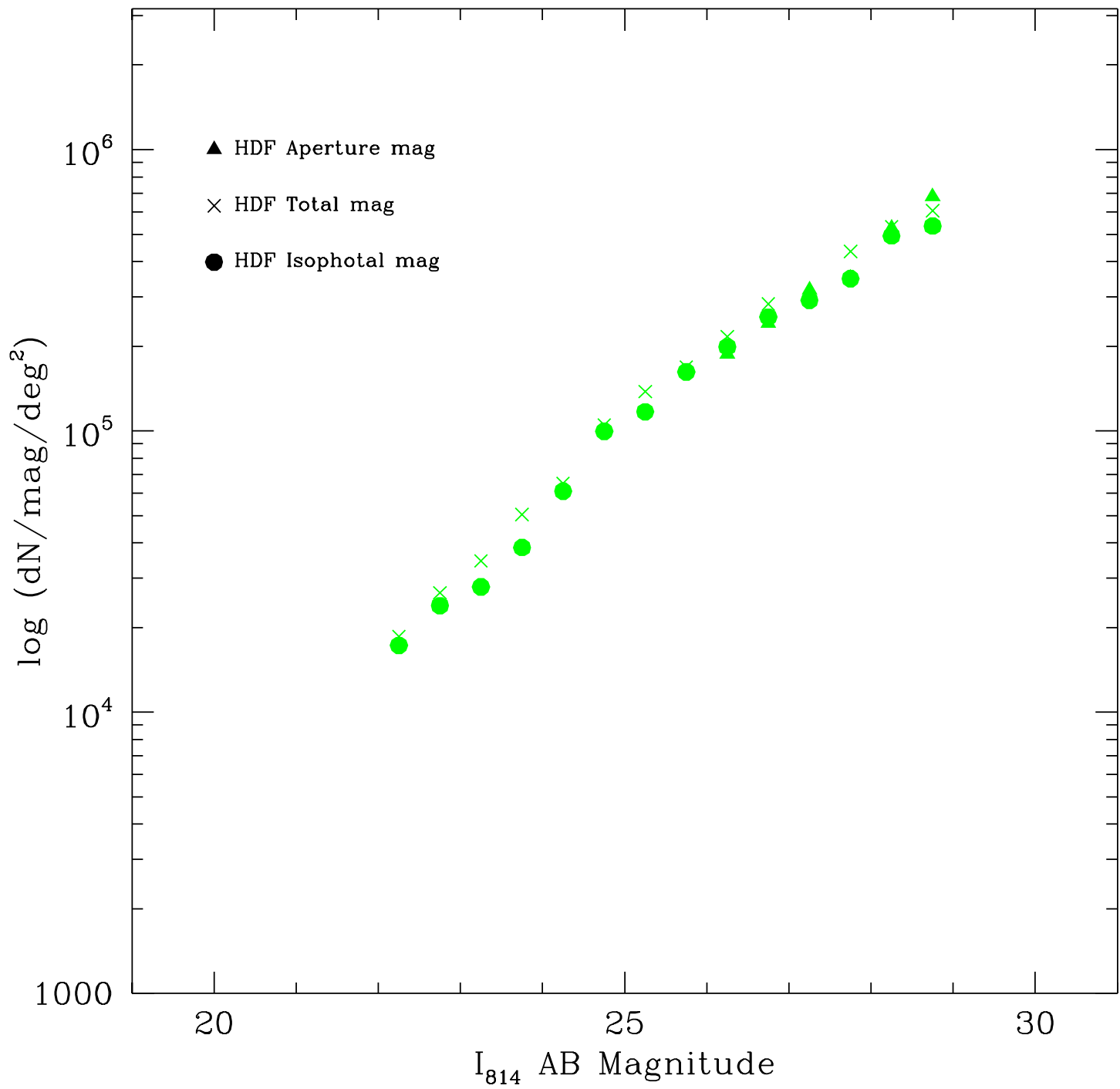
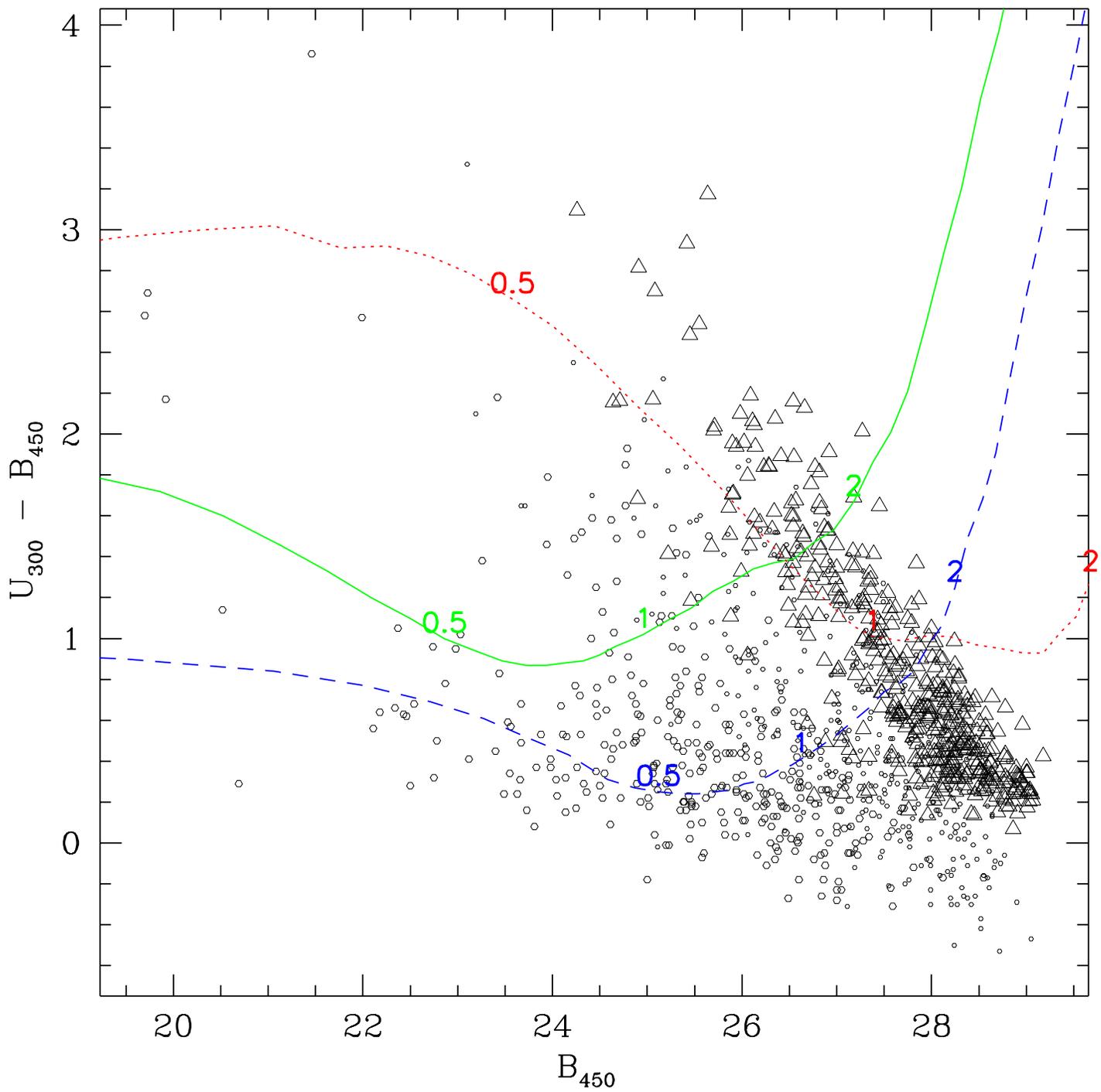
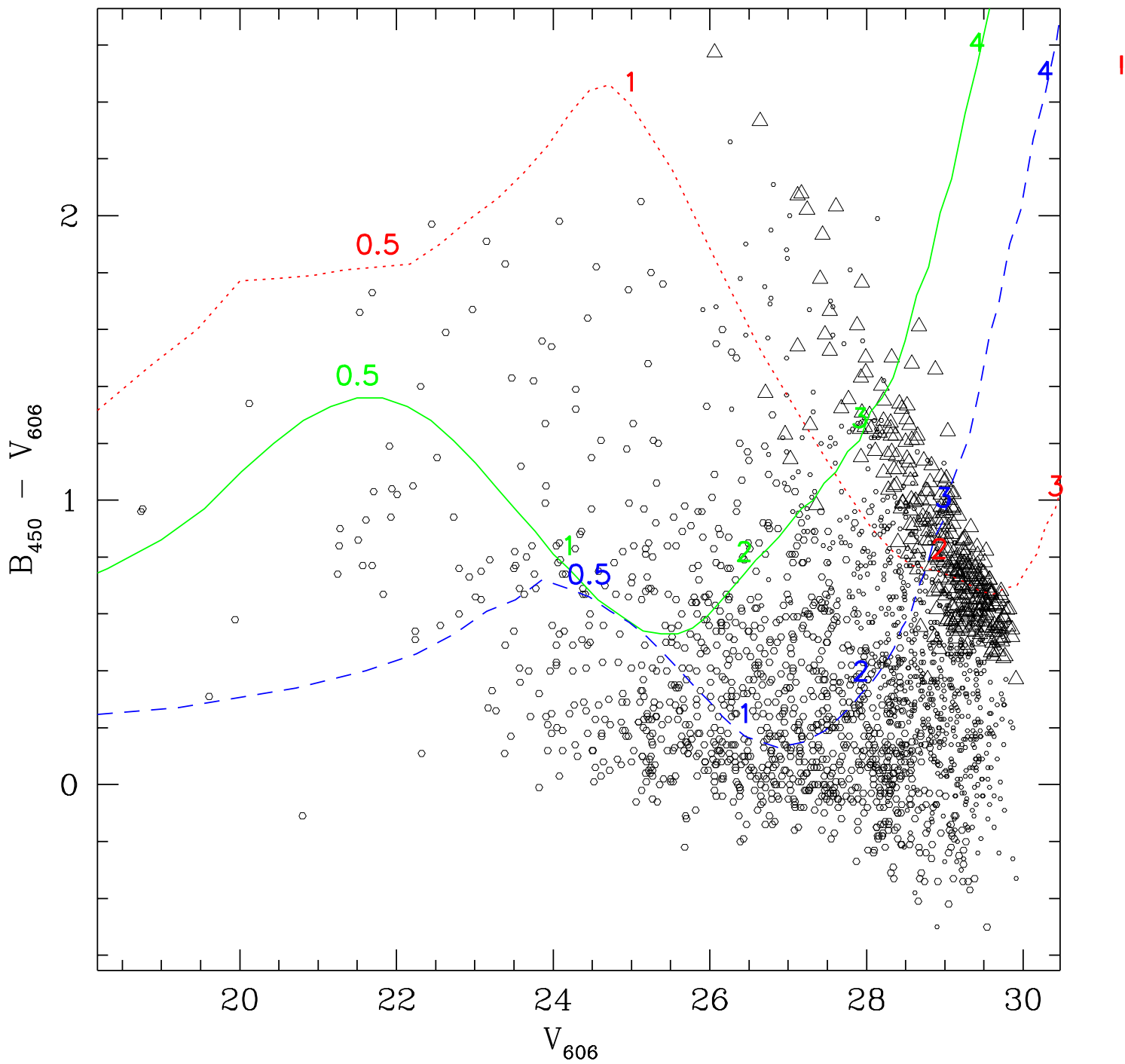
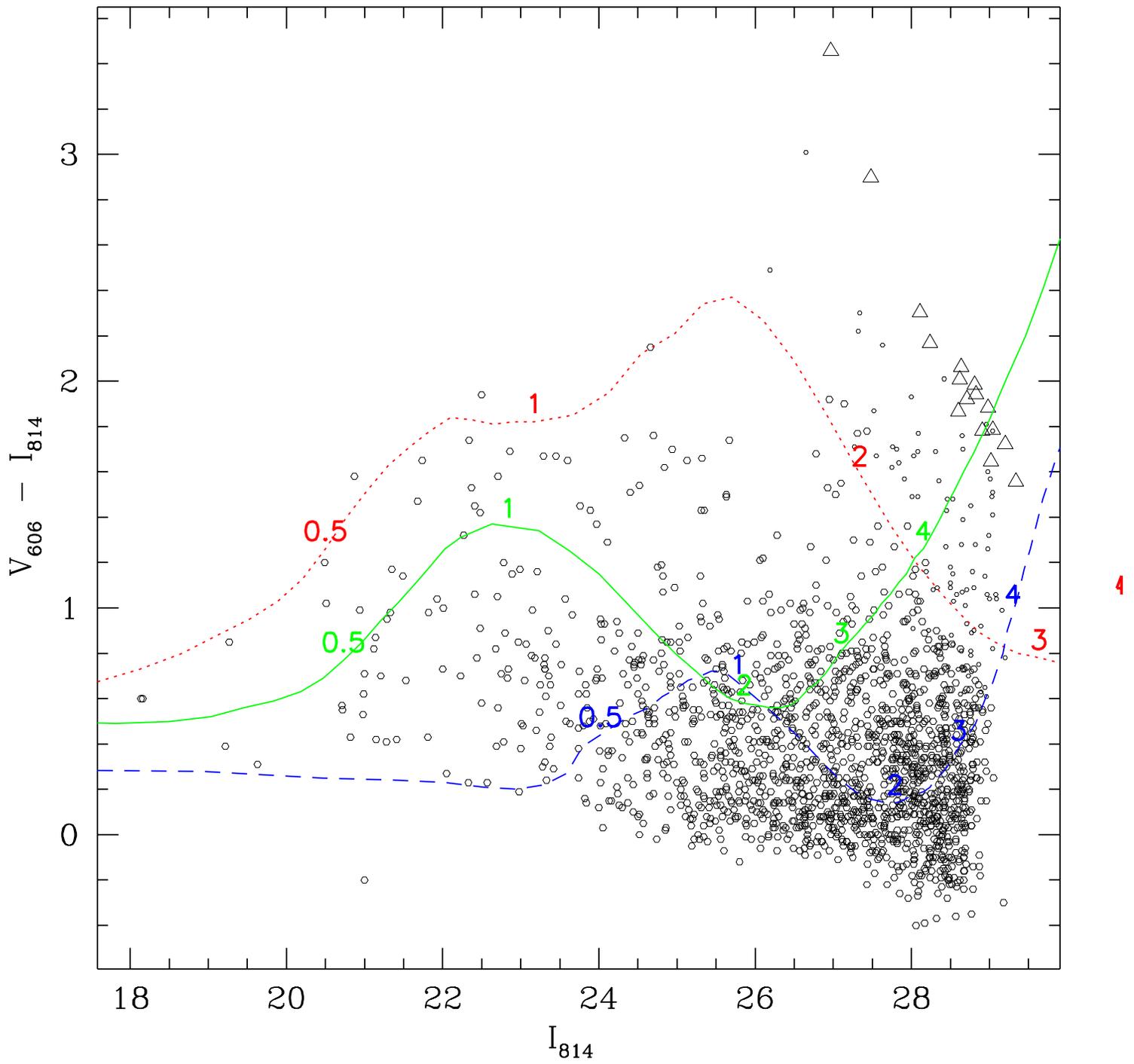
\end{document}